\documentclass[11pt,subeqn,fleqn]{article}
\usepackage{amsfonts,times}
\usepackage{latexsym}
\usepackage{graphicx}

\def\be{\begin{eqnarray}}
\def\ee{\end{eqnarray}}
\def\bmath{\boldmath}
\def\ubmath{\unboldmath}
\def\soso{SO(9)\times SO(9)}
\def\ints{{\mathbb{Z}}}
\def\bi{{\bar \imath}}
\def\bj{{\bar \jmath\,}}
\def\bk{{\bar k}}
\def\bl{{\bar l}}
\def\ba{{\bar\alpha}}
\def\bb{{\bar\beta}}
\def\bg{{\bar\gamma}}
\def\bd{{\bar\delta}}
\def\hm{{\hat m}}
\def\hn{{\hat n}}
\def\hp{{\hat p}}
\def\hq{{\hat q}}
\def\ha{{\hat a}}
\def\hb{{\hat b}}

\def\hG{{\hat{G}}}
\def\sGh{{\sqrt{\hat{G}}}}
\def\sg{{\sqrt{g}}}

\def\G{{\Gamma}}
\def\S{{\Sigma}}
\def\a{{\alpha}}
\def\b{{\beta}}
\def\g{{\gamma}}
\def\d{{\delta}}
\def\s{{\sigma}}
\def\m{{\mu}}

\def\e{{\epsilon}}
\def\ve{\varepsilon}
\def\o{{\omega}}
\def\to{{\tilde \omega}}
\def\L{{\Lambda}}
\def\vp{{\varphi}}
\def\dA{{\dot{A}}}
\def\dB{{\dot{B}}}
\def\dC{{\dot{C}}}

\def\pt{{\partial_t}}
\def\p{{\partial}}
\def\tP{{\tilde{\Psi}}}

\def\cL{{\cal L}}
\def\cD{{\cal D}}

\def\cV{{\cal V}}
\def\cQ{{\cal Q}}
\def\cP{{\cal P}}

\def\boldone{{\bf 1}}
\def\reals{{\mathbb R}}
\def\ints{{\mathbb Z}}
\def\lae{{\mathfrak{e}}}

\def\wG{\widehat{\Gamma}}

\newcommand{\CC}{{\cal C}}
\newcommand{\CCb}{{\bar{\cal C}}}
\newcommand{\wCC}{{\widehat{\cal C}}}
\newcommand{\ft}[2]{{\textstyle {\frac{#1}{#2}} }}

\newcommand{\diag}{{\rm diag \,}}

\newcommand{\nn}{{\nonumber}}
\def\tten{{\widetilde{10}}}
\def\half{\frac{1}{2}}
\def\symbolfootnote[#1]#2{\begingroup%
\def\thefootnote{\fnsymbol{footnote}}\footnote[#1]{#2}\endgroup}

\usepackage{longtable,amsmath,amsfonts}

\begin{document}

\begin{titlepage}
\begin{flushright}
hep-th/0407101\\
AEI-2004-040\\
DAMTP 2004-52\\[25mm]
\end{flushright}
\begin{center}{\LARGE \bf  \bmath $E_{10}$ and $SO(9,9)$ \ubmath invariant
supergravity}\\[18mm]
{\bf A. Kleinschmidt}\symbolfootnote[2]{Address after August 1, 2004:
  Max-Planck-Institut f\"ur Gravitationsphysik, M\"uhlenberg 1,
  D-14476 Golm, Germany}\\
{\sl Department of Applied Mathematics and Theoretical Physics}\\
{\sl Centre for Mathematical Sciences, Wilberforce Road, Cambridge CB3
  0WA, U.K.}\\
a.kleinschmidt@damtp.cam.ac.uk\\[5mm]
{\bf H. Nicolai}\\
{\sl Max-Planck-Institut f\"ur Gravitationsphysik}\\
{\sl M\"uhlenberg 1, D-14476 Golm, Germany}\\
nicolai@aei.mpg.de\\[20mm]
\end{center}

\renewcommand{\abstract}{\begin{center}\bf
    Abstract\\[3mm]\end{center}}

\begin{abstract}

We show that (massive) $D=10$ type IIA supergravity possesses a
hidden rigid $D_9\equiv SO(9,9)$ symmetry and a hidden local
$SO(9)\times SO(9)$ symmetry upon dimensional reduction to one
(time-like) dimension. We explicitly construct the associated
locally supersymmetric Lagrangian in one dimension, and show that
its bosonic sector, including the mass term, can be equivalently
described by a truncation of an $E_{10}/K(E_{10})$ non-linear
$\s$-model to the level $\ell \leq 2$ sector in a decomposition
of $E_{10}$ under its $D_9$ subalgebra. This decomposition is
presented up to level $\ell =10$, and the even and odd level
sectors are identified tentatively with the Neveu--Schwarz and
Ramond sectors, respectively. Further truncation to the level
$\ell =0$ sector yields a model related to the reduction of $D=10$
type~I supergravity. The hyperbolic Kac--Moody algebra $DE_{10}$,
associated to the latter, is shown to be a proper subalgebra of
$E_{10}$, in accord with the embedding of type~I into type IIA
supergravity. The corresponding decomposition of $DE_{10}$ under 
$D_9$ is presented up to level $\ell=5$.

\end{abstract}

\end{titlepage}

\begin{section}{Introduction}

Dimensional reduction of supergravity theories has proved to be
not only a way of constructing lower dimensional theories with
extended supersymmetry, but also a way of revealing
hidden symmetries \cite{CrJu78,Ju81}. The example studied most
thoroughly in this context is that of maximal eleven dimensional
supergravity \cite{CrJuSche78}, which upon dimensional reduction
gives rise to the chain of exceptional hidden symmetries $E_{n(n)}$.
Below three uncompactified dimensions, the relevant Kac--Moody 
algebras are infinite dimensional \cite{Ju82}.

In one dimension, the expected hidden symmetry is the elusive
hyperbolic $E_{10}$ symmetry whose root space geometry is known to
govern the behavior of cosmological solutions of $D=11$
supergravity near a space-like singularity
\cite{DaHe01,DaHeJuNi01,DaHeNi03}. These so-called cosmological billiards
were also studied in the context of other over-extended Kac--Moody
algebras, see
\cite{DadeBuHeScho02,HeJu03,deBuPiScho03,deBuHeJuPa03,deBuScho04}.
In the toroidal compactification of $D=11$ supergravity
to one time dimension there is a manifest
$GL(10;\reals)$ symmetry acting on the internal components. By
making use of a `level expansion' of $E_{10}$ in terms of
$GL(10;\reals)$ tensors, it was shown in \cite{DaHeNi02} that the
equations of motion of a $\s$-model on the $E_{10}/K(E_{10})$
coset space, when truncated to the first three levels, are
equivalent to a restricted version of the bosonic equations of
motion of $D=11$ supergravity where only the fields and their
first order spatial gradients at a given spatial point are
retained. It was furthermore shown there that the level
decomposition of $E_{10}$ contains representations that can be
naturally associated to the spatial gradients of the bosonic
$D=11$ supergravity fields. This observation gave rise to the
conjecture that the geodesic motion on the $E_{10}/K(E_{10})$
coset manifold can capture the full space-time dependence of the
$D=11$ supergravity fields, such that the standard BKL
approximation is recovered via a `small tension expansion' in
spatial gradients. However, the  $E_{10}$ $\sigma$-model is
expected to contain many further degrees of freedom, because the
representations corresponding to spatial gradients constitute only
a tiny subset of the $E_{10}$ Lie algebra \cite{NiFi03}.

The aim of the present paper is two-fold. First, we demonstrate
that (massive) type IIA supergravity \cite{Ro86} reduced to one
time-like dimension admits a hidden symmetry $SO(9,9)/\soso$.
The same symmetry appears already for the reduction of
type I supergravity to one dimension, which is obtained from the
IIA theory by restricting to the Neveu--Schwarz type fields. These
results extend earlier ones on the emergence of $SO(n,n)$
symmetries in dimensionally reduced type I supergravity
\cite{Cha81,MaSchw93,SchSen93}. The hidden $SO(9,9)$ symmetry
group has a natural interpretation in string theory, being the
continuous T-duality group of the low energy effective action of
the two type II string theories reduced to one dimension, whose
discrete subgroup $SO(9,9;\ints)$ is known to be a good quantum
symmetry of the perturbative string spectrum \cite{Na86,NaSaWi87}.

The second main  result of this paper is an analysis of the
$E_{10}/K(E_{10})$ coset model at low levels in terms of a level
decomposition of $E_{10}$ under its $D_9 \equiv SO(9,9)$ subgroup,
complementing the results of \cite{DaHeNi02} where a level
decomposition of $E_{10}$ w.r.t. its $A_9 \equiv SL(10;\reals)$
subgroup was used. Just like $SL(10,\reals)$, the group $SO(9,9)$ is a
regular subgroup of $E_{10}$, and its maximal compact subgroup
$\soso$ is a subgroup of $K(E_{10})$, the maximal compact subgroup
of $E_{10}$. Accordingly, we now decompose $E_{10}$ into an
infinite tower of $SO(9,9)$ representations, with the level as the
`floor number'. We then proceed to use this decomposition to
compare the dynamics of the truncated $\s$-model to the reduced
bosonic equations of motion.

In comparison with \cite{DaHeNi02}, the present results reveal
several new facets. In particular, the decomposition of $E_{10}$
into $SO(9,9)$ assigns a special role to the $10$-th spatial
dimension. The compact $\soso$ subgroup has as its diagonal
subgroup $SO(9)_\diag$, which is the remnant of the Lorentz group
acting on nine compact spatial dimensions only. The last
coordinate direction corresponds to a dilatonic field $\vp$ (not
the standard IIA dilaton), which is orthogonal to $SO(9,9)$ and
algebraically provides the desired grading of $E_{10}$ w.r.t. its
$D_9$ subalgebra. Another new feature is the appearance of {\em
spinorial representations} of $SO(9,9)$ at odd levels in the level
decomposition: while the vectorial representations corresponding
to the Neveu--Schwarz-Neveu--Schwarz (NSNS) fields appear at {\em
even} levels, the Ramond-Ramond (RR) type fields are associated
with the {\em odd} levels. Our terminology here is loose in the
sense that we refer to any state composed as a product of two
vectorial representations of $SO(9)$ as an NSNS field, and to any
state composed out of two spinorial representations of $SO(9)$ as
an RR field. Restricting to the even level NSNS sector, one
obtains the non-maximal type I theory which is simpler to study in
the present context. Its reduction to one dimension likewise
admits a $SO(9,9)/\soso$ invariant formulation. In the
corresponding $E_{10}$ model, the type I theory corresponds to
level $\ell=0$ in the decomposition which forms a closed
subalgebra of $\lae_{10}$. Because $SO(9,9)$ is also contained in
$DE_{10}$, which has been conjectured to be a hidden symmetry of
type I supergravity reduced to one dimension \cite{Ju82,DaHe01},
analogous statements hold true for the coset $DE_{10}/K(DE_{10})$.
The mutual consistency of these conjectures is a consequence of
the fact that $DE_{10}$ is 
generated by a proper (irregular) subalgebra of
$\lae_{10}$. When extending these results to type IIA supergravity we
have to embed RR fields into representations of $SO(9,9)$, which
we are able to do by adding their duals and a Romans mass term
(which in the present context is best understood in terms of a
9-form \cite{Ro86,BedeRoGrPaTo96,ObPiRa98,ObPi99}). This is in
precise agreement with the structure found at $\ell=1$ in the
decomposition of $E_{10}$ under its $D_9$ subalgebra, and shows
that {\em massive} IIA supergravity, too, can be accommodated
within $E_{10}$.

Unlike previous work, our analysis also includes the fermionic degrees
of freedom, albeit only `at level $\ell =0$'. In particular, we
demonstrate that the fermions can also be placed into multiplets of
$\soso$ in this approach. Furthermore, we determine the relevant
$\sigma$-model quantities from the supersymmetry variations; the bosonic 
equations of motion, which were used for this purpose in \cite{DaHeNi02},
then provide an additional and independent consistency check.
While the supersymmetry algebra in the type I case closes on 
shell on the finite set of fields obtained by reduction
of type I supergravity, it acquires new terms in the type II case,
which no longer close on the finite set of fields obtained by
$\soso$ `covariantization' of the type II supergravity fields. The
necessity of (in fact, infinitely many) new fields follows
algebraically by noting that, unlike the level 0 representations,
the level $\ell=-1,0,1$ representations do not form a subalgebra
of $\lae_{10}$, but instead generate all of the $\lae_{10}$
algebra. However, it must be stressed that even the most basic
aspects of the fermionic sector in relation to the hyperbolic
symmetry remain to be understood. Whereas it is known at least in
principle how to recursively construct the additional bosonic
degrees of freedom in the $E_{10}$ model (cf. the tables of
\cite{NiFi03} and appendices B and C of this paper), it remains
an outstanding challenge to also extend the fermionic $\soso$
multiplets to a full spinor (i.e. double valued) representation
of $K(E_{10})$, and to write down a locally supersymmetric model
compatible with local $K(E_{10})$ symmetry\footnote{See, however,
\cite{NicSam98,NicSam04} for some recent results on the (much
simpler) involutory $K(E_9)$ symmetry of the $d=2$ theory.}.

A non-linear realization of massive IIA supergravity \cite{SchnWe02},
and the embedding of the bosonic sector of type I into that of
type II supergravity \cite{SchnWe04} were already investigated in
the context of an earlier, and conceptually different, proposal
concerning the realization of hidden Kac--Moody symmetries
in M-theory \cite{We01,We03a,We03b,We04} (see also
\cite{KlSchnWe04,KlWe04,SchnWe01}). According to that proposal,
$D=11$ supergravity admits an M-theoretic extension possessing
an even larger symmetry containing a hidden $E_{11}$, 
which is also supposed to accommodate type IIB supergravity
\cite{SchnWe01}.  
The indefinite Kac--Moody algebra $E_{11}$ is not hyperbolic 
any more, but belongs to a class of Lorentzian Kac--Moody algebras 
called very extended algebras \cite{GaOlWe02}. In a related development, 
the Weyl group of $E_{11}$ was shown to act on the moduli of 
Kasner solutions \cite{EnHoTaWe03} and to generate the intersection 
rules for branes \cite{EnHoWe03}. A $\sigma$-model approach,
which aims to merge the proposals of \cite{DaHeNi02} and \cite{We01}
by introducing an unphysical auxiliary parameter, was recently
formulated and studied in \cite{EnHo03,EnHo04b}.

For $E_{10}$, a `brany' interpretation of the imaginary roots has been
proposed in \cite{BrGaHe04}; furthermore, these authors conjecture
that essential information about M-theory compactified on $T^{10}$
is contained in a generalized modular form over the coset space
$E_{10}(\ints) \!\setminus\! E_{10}/K(E_{10})$. This modular form can
be viewed as the solution of the Wheeler--DeWitt equation that is
(formally) obtained by quantizing the Hamiltonian constraint of
\cite{DaHeNi02}. Very similar ideas have been put forth in
\cite{PioWal02}. Finally, an attempt to merge M(atrix) theory
and $E_{11}$ has been made in \cite{Cha04}.

The paper is organized as follows. We first explain our conventions
and notations in section~2. In section \ref{so99sec} the
$SO(9,9)$ invariance of reduced type I and type IIA supergravity
is studied indicating the necessity of introducing additional
fields to obtain an $SO(9,9)$ invariant, locally supersymmetric
theory in the latter case.
This could be provided for by studying the $E_{10}$ model
and we carry out the required decomposition in section
\ref{e10rels}. These results are then exploited in section
\ref{sigmamod} to give the equations of motion of the truncated
$\s$-model based on $E_{10}$ where the resulting dynamics are
linked to the reduced supergravities from section \ref{so99sec}.
In appendix \ref{gconv}, we fix our choice of gamma matrices.
Appendices \ref{eddec} and \ref{deddec} contain the details of the
decomposition of $E_{10}$ and $DE_{10}$ with respect to their
regular $D_9$ subalgebras. This decompositions are presented up to
levels $\ell=10$ for $E_{10}$ and $\ell=5$ for $DE_{10}$ in this
paper.\footnote{The result is known up to levels $\ell=20$ for
  $E_{10}$ and up to $\ell=8$ for $DE_{10}$, respectively. The
  relevant data can be obtained through the source file on the
  preprint arXiv.} 

\end{section}

\begin{section}{\bf Conventions}
\label{Conventions}
\setcounter{equation}{0}

The action of $D=11$, $N=1$ supergravity to second order in
fermions \cite{CrJuSche78} in our conventions is
\be
\label{maxsugra}
S&=&\int d^{11}x E
\left(-\frac{1}{4\kappa_{11}^2}R
-\frac{i}{2}\bar\Psi_M\Gamma^{MNP}\nabla_N\Psi_P
-\frac{1}{48}F_{MNPQ}F^{MNPQ}\right)\nn\\
&&- \frac{i \kappa_{11}}{96} \int d^{11}x E \left(\bar\Psi_M\G^{MNPQRS}\Psi_S
+12\bar\Psi^N\G^{PQ}\Psi^R\right)F_{NPQR}\\
&& + \frac{2\kappa_{11}}{(144)^2} \int d^{11}x
\e^{M_1\ldots M_{11}}F_{M_1\ldots M_4} F_{M_5\ldots M_8} A_{M_9\ldots
  M_{11}}
+\ldots.\nn
\ee
We will set Newton's constant to $\kappa_{11}=1$. The ellipsis denotes
additional couplings and higher order Fermi terms which will not be of
relevance in our analysis. The above action is invariant under the 
supersymmetry variations \cite{CrJuSche78}
\begin{subequations}\label{susyvars}
\be
\label{vsusy}
\d {E_M}^A &=& i \bar\e \G^A \Psi_M,\\
\label{gsusy}
\d A_{MNP} &=& \frac{3}{2}i\bar\e\G_{[MN}\Psi_{P]},\\
\label{fsusy}
\d\Psi_M &=& \left(\partial_M \e + \frac{1}{4}\omega_{M\,AB}\G^{AB}\e\right)
-\frac1{144} \big( {\G_M}^{NPQR} - 8\d_M^{N} \G^{PQR}\big)\e
F_{NPQR},
\ee
\end{subequations}
where we have only kept terms up to linear order in fermions in the
variation of the fermions.

From (\ref{maxsugra}) we deduce the bosonic equations of motions 
\begin{subequations}\label{sugraeom}
\be\label{gaugeeom}
\partial_M\big(E F^{MNPQ}\big) &=& -\frac{1}{576}E
\e^{NPQR_1\ldots R_4S_1\ldots S_4}F_{R_1\ldots R_4}F_{S_1\ldots
S_4},\\\label{einsteineom}
R_{MN} &=& -\frac13 F_{MP_1P_2P_3}F_{N}{}^{P_1P_2P_3}
  +\frac1{36}G_{MN}F_{P_1P_2P_3P_4}F^{P_1P_2P_3P_4}.
\ee
\end{subequations}
In addition we have the Bianchi identity
\be\label{Bianchi}
\partial_{[M} F_{NPQR]} = 0.
\ee

Let us also fix our notations for the reduction from eleven to one
dimension at this point. We perform a $(1+9+1)$ split of the coordinates 
and label the coordinates of the three different sectors according 
to the following list:
\be
\begin{array}{rclll}
M,N,... &=& t,\tilde{1},...,\tten
        &&  \mbox{curved indices in $D=11$} \\
A,B,... &=&  0,1,...,10  && \mbox{flat indices in $D=11$} \\
\hm,\hn,... &=& \tilde{1},...,\tten
        && \mbox{curved spatial indices in $D=11$} \\
\ha,\hb,... &=& 1,...,10 && \mbox{flat spatial indices in $D=11$}\\
m,n,... &=& \tilde{1},...,\tilde{9}
         && \mbox{curved spatial indices in $D=10$}\\
a,b,... &=& 1,...,9 && \mbox{flat spatial indices in $D=10$}\end{array}\nn
\ee
To distinguish flat indices from curved ones, we will put
a tilde on the latter, e.g. $\tten$. Below, we will furthermore need
indices pertaining to the $SO(9,9)/\soso$ coset space;
these are
\be
\begin{array}{rclll}
I,J,...&\in& \{1,...,18\} &&\mbox{for $SO(9,9)$}\\
i,j,...&\in& \{1,...,9\} && \mbox{for the first $SO(9)$}\\
\bi,\bj,...&\in& \{\bar{1},...,\bar{9}\} \equiv \{10,...,18\}
    && \mbox{for the second $SO(9)$}\\
\a,\b,...&\in& \{1,...,16\}
    && \mbox{spinor indices for the first $SO(9)$}\\
\ba,\bb,...&\in& \{1,...,16\}
    && \mbox{spinor indices for the second $SO(9)$}\end{array}\nn
\ee
Finally, in later sections, we will also use indices $A,B,...$
for the 256-dimensional chiral spinor representations of $SO(9,9)$ and
$\dA,\dB,...$ for the conjugate spinor. Our $\g$-matrix conventions
for the orthogonal groups $SO(9)$, $SO(1,10)$ and $SO(9,9)$ are
listed in the separate appendix \ref{gconv}.

\end{section}

\begin{section}{\bf \bmath $SO(9,9)$ \ubmath
 invariant supergravity in one dimension}
\label{so99sec}
\setcounter{equation}{0}

In this section we study the dimensional reduction of (\ref{maxsugra})
to one timelike dimension, but in
a setting that is appropriate to IIA supergravity, and the $SO(9,9)$
symmetry which we wish to exhibit. To this end we perform a (9+1) split
of the ten spatial coordinates taking us through $D=10$ supergravity.
We first study the type I theory and find that it can be written as a
coset model on the space $SO(9,9)/\soso$ with additional
dilaton. Extending the analysis to type IIA supergravity the
additional fields generate new terms both in the Lagrangian and the
supersymmetry variations and a locally supersymmetric version as a
coset model is no longer possible with a finite dimensional coset space.

We first consider the reduction of the elfbein degrees of freedom in
the $(1+9+1)$ split of the coordinates. Keeping an arbitrary
lapse function $N$ and setting the shift variables to zero,
the elfbein of $D=11$ supergravity reads
\be
{E_M}^A = \left(\begin{array}{cc} N
&0\\0& {\hat{E}_\hm}{}^{\ha} \end{array}\right)
\quad\Longrightarrow\quad G_{MN}= \left(\begin{array}{cc} -N^2
&0\\0& \hat{G}_{\hm\hn}
\end{array}\right).
\ee
$N$ is a Lagrange multiplier enforcing reparametrisation
invariance w.r.t. to the diffeomorphisms generated by timelike
vector fields. Next we perform a (9+1) split of the spatial part
of the elfbein in triangular gauge:
\be\label{spatialvsplit}
{\hat{E}_\hm}{}^{\ha}=\left(\begin{array}{cc}{E_m}^a &
e^{\frac12\phi} A_m
\\0&e^{\frac12 \phi} \end{array}\right) \quad\Longrightarrow\quad
{\hat{E}_\ha}{}^{\hm}=\left(\begin{array}{cc}{E_a}^m &
- {E_a}^m A_m
\\0&e^{-\frac12 \phi} \end{array}\right).
\ee
The spatial metric and its inverse are given by
\be
\hat{G}_{\hm\hn}=\left(\begin{array}{cc}G_{mn} + e^\phi A_m A_n& e^\phi A_m
\\ e^\phi A_n &e^\phi
\end{array}\right)
\quad\Longrightarrow\quad
\hat{G}^{\hm\hn}=\left(\begin{array}{cc}G^{mn}& - A^m \\
- A^n & e^{-\phi} + A^r A_r \end{array}\right).
\ee
Here, the index on $A^m$ has been raised with $G^{mn}$. In order to
end up with a Lagrangian that has the required symmetries, we must
still redefine the metric $G_{mn}$ and the dilaton $\phi$. The properly
redefined fields $g_{mn}$ and $\vp$ of the reduced theory are
\be
\label{nscos1}
e^\vp =  G^{-\frac{1}{2}} e^{-\frac{3}{4}\phi} \quad , \qquad
{e_m}^a = e^{\frac14 \phi} {E_m}^a \quad \Longrightarrow \quad
g_{mn} = e^{\frac{1}{2}\phi} G_{mn}.
\ee
By $G$ and $g$, we denote the determinants of $G_{mn}$ and $g_{mn}$,
respectively; thus $g= e^{\frac92\phi}G$. Inverting these relations,
we obtain
\be
G_{mn} = g^{\frac{1}{6}}e^{-\frac{1}{3}\vp}g_{mn}\quad , \qquad
e^\phi = g^{\frac{1}{3}}e^{\frac{2}{3}\vp}.
\ee
While the field $\phi$ thus is just the standard dilaton, our
`dilaton' $\vp$ differs from it by its additional dependence on the
9-metric $g_{mn}$. For this reason, $\vp$ will appear in the
Lagrangian in a way different from the way in which $\phi$ appears
in the standard type I and type IIA Lagrangians.

A short calculation now shows that the reduction of Einstein's action to
one dimension indeed becomes diagonal in terms of the fields $g$, $\vp$
and $A_m$, viz.
\be\label{redeinstein}
\hat{G}^{\hm\hn}\pt{\hat{G}}_{\hn\hp} \hat{G}^{\hp\hq}
\pt{\hat{G}}_{\hat\hq\hm}
-\left(\hat{G}^{\hm\hn}\pt{\hat{G}}_{\hm\hn}\right)^2
&=&g^{mn}\pt{g}_{np}g^{pq}\pt{g}_{qn} -4\pt\vp \pt\vp    \nn\\
&&+2e^{\vp}g^{\frac12} g^{mn} \pt{A}_m \pt{A}_n.
\ee

Next, we turn to reduction of the eleven dimensional gravitino $\Psi_M$.
Its temporal component $\Psi_t$, with 32 real spinor components,
is a Lagrange multiplier field (enforcing the supersymmetry constraint),
while its spatial components constitute a vector spinor of $SO(10)$
with a total of $10\times 32 = 320$ real components, corresponding to
the $\bf{32}$ and $\bf{288}$ representations of $SO(10)$.
A 32-component  Majorana spinor in eleven dimensions reduces to two
16-component real spinors of $SO(9)$, while the vector part reduces to
a vector plus a scalar. The subscript $t$ is to indicate the transformation
properties of $\Psi_t$ under reparametrizations of the time
coordinate $t$; the corresponding `flat' object is $\Psi_0=N^{-1}\Psi_t$.
With this split of the coordinates we introduce redefined fermion
fields as
\begin{subequations}\label{Psitilde}
\be
\tP_t &=& \hG^{-\frac{1}{4}}
          \Big(\Psi_t-\G_t\G^a\Psi_a-\G_t\G^{10}\Psi_{10}\Big),  \\
\tP_a &=& \hG^{\frac{1}{4}}
           \Big(\Psi_a+\frac{1}{2}\G_a\G^{10}\Psi_{10}\Big),   \\
\tP_{10} &=& \hG^{\frac{1}{4}}
           \Big(-\frac{3}{2}\Psi_{10}-\G_{10}\G^{a}\Psi_a\Big),
\ee
\end{subequations}
with $\hG=\det \hG_{\hm\hn} = G e^\phi$. Inverting these formula, we obtain
\begin{subequations}\label{Psiinv}
\be
\Psi_t &=&
\hG^{\frac{1}{4}}\tP_t
    -\frac{1}{6}\hG^{-\frac{1}{4}}\G_t\G^a\tP_a
    -\frac{7}{6}\hG^{-\frac{1}{4}}\G_t\G^{10}\tP_{10},  \\
\Psi_a &=&  \hG^{-\frac{1}{4}}
     \Big(\tP_a-\frac{1}{6}\G_a\G^b\tP_b-\frac{1}{6}\G_a\G^{10}\tP_{10}\Big),\\
\Psi_{10} &=& \hG^{-\frac{1}{4}}
     \Big(\frac{1}{3}\tP_{10}+\frac{1}{3}\G_{10}\G^a\tP_a\Big).
\ee
\end{subequations}
With the above redefinitions one can check that the derivative
part of the fermion kinetic term reduces as
\be\label{kinfermion}
-\frac{i}{2}N\sGh \bar{\Psi}_A\G^{ABC}\partial_B\Psi_C
=\frac{i}{2}\tP_a^T\pt\tP_a-\frac{i}{2}\tP_{10}^T\pt\tP_{10},
\ee
where we have used that $\bar\Psi=\Psi^T\G^0$ in our conventions.
The connection terms (of type $\omega \Psi^2$) in the kinetic term
will be shown below to combine with the $F \Psi^2$ terms in (\ref{maxsugra})
to give an $\soso$-covariantization of the derivative above.

In the supersymmetry variations, the above redefinitions must be
accompanied by the following rescaling of the supersymmetry
transformation parameter with the opposite power of $\hG$
\be\label{SUSYparameter}
\e \longrightarrow \ve \equiv \hG^{-1/4} \e.
\ee
It is noteworthy that the redefinitions of the fermionic 
variables are the ones that one would expect on account of the
corresponding formulae in higher dimensions; for instance, the
general formula for the redefinitions of the gravitino field
and the supersymmetry transformations parameter in Kaluza--Klein
supergravity read
\be
\psi'_\mu = \hG^{\frac14 \frac1{2-d}} {e'_\mu}^\alpha
\big(\Psi_\alpha + \frac1{d-2} \Gamma_\alpha \Gamma^a \Psi_a\big)
\quad , \qquad
\ve' = \hG^{-\frac14 \frac1{2-d}} \ve
\ee
where $\alpha$ and $a$ label the (flat) uncompactified and compactified
coordinates, respectively, $d$ is the number of uncompactified
dimensions, and $\hG$ the determinant of the compactified part of
the metric. Although these formulae obviously fail for $d=2$ they
do work again for $d=1$!

\begin{subsection}{\bf Type I theory}

We first restrict to the case of type I supergravity and show that it
can be written with local $\soso$ invariance in both the bosonic and
the fermionic sector.

The bosonic fields of this theory are the zehnbein, an antisymmetric
2-form field and a dilaton, referred to as Neveu--Schwarz-Neveu--Schwarz
fields in IIA string theory. Therefore we set the Kaluza--Klein vector of
the metric $A_m=0$ in this section.
The 2-form field of type I supergravity originates from the 3-form
field $A_{MNP}$ in eleven dimensions, with $P=\tten$; all other
components of the 3-form will be set to zero for the type I theory.
The corresponding terms are most conveniently derived by first
writing the relevant term in eleven dimensions with flat indices,
and then reconverting to curved indices by means of the appropriate
rescaled vielbeine. For the field $b_{mn}=2 A_{mn\tten}$ and adopting
the Coulomb gauge $A_{tmn} =0$, this procedure yields for the kinetic term
\be\label{FF1}
\frac{1}{12}N \sGh \hG^{tt} G^{\hm\hq} G^{\hn\hat{r}} G^{\hp\hat{s}}
   F_{t\, \hm\hn\hp} F_{t\, \hq\hat{r}\hat{s}} &\rightarrow& \nn\\
-\frac{1}{4}N^{-1} \sGh \hG^{mp} \hG^{nq} \hG^{\tilde{10}\tilde{10}}
   F_{t\, mn\tilde{10}} F_{t\, pq\tilde{10}} &=&-\frac{1}{16}
N^{-1} \sqrt{\hat{G}} g^{mp} g^{nq} \pt b_{mn} \pt b_{pq};
\ee
note that the dilaton factor in the inverse metric prefactors
cancels by virtue of our redefinition (\ref{nscos1}). Combining this
with the result for the reduced Einstein action (\ref{redeinstein})
the reduced action for the type I theory therefore reads
\be\label{typeI}
\frac{1}{16}\int dt\, N^{-1}  \sGh
\left(g^{mn}\pt{g}_{np}g^{pq}\pt{g}_{qm}
-g^{mn}\pt{b}_{np}g^{pq}\pt{b}_{qm} - 4\pt\vp \pt\vp\right).
\ee
In string frame the type I bosonic Lagrangian comes with a dilatonic
prefactor \cite{Po98}. Here, however, we do not want such a factor, because
we wish to embed this Lagrangian into an $E_{10}/K(E_{10})$ $\s$-model
such that the level $\ell$ terms appear with a factor $e^{\ell\vp}$,
and therefore there should be no dilatonic prefactor for $\ell=0$
Lagrangian, corresponding to type I. For this reason we redefine
the lapse according to
\be
N=  \sqrt{\hat{G}} \cdot n
\ee
with the new lapse function $n(t)$ (the gauge used in the cosmological 
billiard description \cite{DaHeNi03} is then $n=1$).
The supersymmetry partner of the latter is the redefined gravitino
\be
\tP_t \equiv N \tP_0 = n\hG^{\frac{1}{4}}
          \Big(\Psi_0-\G_0\G^a\Psi_a-\G_0\G^{10}\Psi_{10}\Big)
\ee

The bosonic type I Lagrangian (\ref{typeI}) possesses a hidden $SO(9,9)$
symmetry which we can exhibit explicitly by parametrising the fields
$g$ and $b$ in terms of a representative of the coset space
${\cal E}\in SO(9,9)/\soso$
\be
\label{calE}
{\cal  E}(t)=\frac{1}{2}\left(\begin{array}{cc}1&1\\-1&1\end{array}\right)
\left(\begin{array}{cc}e^{-1}&-e^{-1}b\\0&e^T
\end{array}\right)\left(\begin{array}{cc}1&-1\\1&1\end{array}\right)
\ee
in self-explanatory matrix notation (the calculation is analogous
to the one in \cite{MaSchw93}). Because there are now {\em two}
$SO(9)$ groups, we will use unbarred and barred $SO(9)$ indices $i,j,...$
and $\bi,\bj,...$, respectively, to distinguish them; the previous $SO(9)$
indices $a,b,...$ then refer to the {\em diagonal subgroup} of $\soso$.
The metric $g_{mn}=e_m{}^a e_n{}^a$ can be read off from
\be
{\cal E}^T{\cal E}=\frac{1}{2}
\left(\begin{array}{cc}1&1\\-1&1\end{array}\right)
\left(\begin{array}{cc}g^{-1}&-g^{-1}b\\bg^{-1}&
g-bg^{-1}b\end{array}\right)
\left(\begin{array}{cc}1&-1\\1&1\end{array}\right).
\ee
To write down the Lagrangian, we need
\be
\label{cosetfield}
 \pt{\cal E}{\cal E}^{-1} = \ft{1}{2}Q^{ij}X_{ij}+
\ft{1}{2}Q_{\bi\bj}X^{\bi\bj}+P_{i\bj}Y^{i\bj}.
\ee
where $(Q_{ij},Q_{\bi\bj})$ is the gauge field for the compact $\soso$
subgroup. The Lie algebra elements $X^{ij}$ and $X^{\bi\bj}$ are
the generators of the compact
$\soso$ subgroup and the $Y^{i\bj}$
are the remaining (broken) generators of the full $SO(9,9)$.
The explicit expressions are
\begin{subequations}\label{QP}
\be\label{Q}
Q_{ij} &=& { e_{[i}}^m \pt e_{m j]} +\half
           {e_{i}}^m{e_{j}}^n \pt b_{mn} ,\\
\label{Qbar}
Q_{\bi\bj} &=& {e_{[i}}^m \pt e_{m j]} - \half
           {e_{i}}^m{e_{j}}^n \pt b_{mn} ,\\
\label{P}
P_{i\bj} &=&   {e_{(i}}^m \pt e_{m j)} - \half
           {e_{i}}^m {e_{j}}^n \pt b_{mn}.
\ee
\end{subequations}
Not forgetting the dilaton, which transforms as a singlet under $SO(9,9)$,
and whose kinetic term appears with a minus sign, the bosonic Lagrangian from
equation (\ref{typeI}) can be written as
\be
{\cal L}_I = \frac{1}{4}n^{-1} \big( P_{i\bj} P_{i\bj}
        - \pt\vp \pt\vp \big)
\ee
where $n$ is a lapse function already introduced above.

To exhibit the $\soso$ invariance of the fermionic sector, we regroup
the 320 gravitino components into irreducible multiplets of $\soso$.
Because a 32-component Majorana spinor in $D=11$ reduces to two
16-component $D=10$ spinors and a $D=10$ vector reduces to a $D=9$
vector and a scalar, we split the spatial components of the $D=11$
gravitino as follows:
\be\label{IIAfermions}
\begin{array}{ccccccccc}{\bf 320}\rightarrow&
({\bf 9},{\bf 16})&+&({\bf 1},{\bf 16})
&+& ({\bf 16},{\bf 9})&+&({\bf 16},{\bf 1}) \\
&\chi_{i\ba}&&\chi_{\ba}&&\chi_{\bi\a}&&\chi_{\a}
\end{array}
\ee
The essential point here is that the vector and spinor indices
are now decreed to transform under {\em two different} $SO(9)$
groups as indicated by the replacement of $a$ by $i$ and $\bi$,
respectively, and the split of $SO(1,10)$ spinor indices into $\a$ and $\ba$.
The consistency of these assignments does not follow from dimensional
reduction alone, but must instead be verified by direct computation.
The kinetic term for the fermions from equation (\ref{kinfermion})
can be written as
\be
\label{fermkin}
\frac{i}{2}\tilde\Psi_a^T\partial_t\tilde\Psi_a
  -  \frac{i}{2}\tilde\Psi^T\partial_t\tilde\Psi
= \frac{i}{2}\chi_{i\ba}\pt\chi_{i\ba}
+\frac{i}{2}\chi_{\bi\a}\pt\chi_{\bi\a}
-  \frac{i}{2}\chi_\a\pt\chi_\a -  \frac{i}{2}\chi_\ba\pt\chi_\ba,
\ee
exhibiting global $\soso$ invariance \footnote{We note that this part 
of the Lagrangian actually has a hidden $SO(1,9) \times SO(1,9)$ invariance, 
which however does not seem to extend to the full Lagrangian.
The non-compact form is essential here because the $\soso$ 
assignments for the fermionic fields (\ref{IIAfermions})
cannot be extended to $SO(10)\times SO(10)$ representations
without doubling the number of fermionic components: the latter
would necessarily belong to the $(\bf{10},\bf{32}) \oplus
(\bf{32},\bf{10})$ representation of $SO(10)\times SO(10)$.
Killing equations for doubled fermions were derived recently
in \cite{MiSchn04}.}.
The derivative terms in (\ref{fermkin}) combine with the
contributions from the spin connection and $F \Psi^2$ terms in
the $D=11$ supergravity theory in such a way as to give a
covariantization of the derivative under local $\soso$
transformations. As we are considering the type I theory in this
section, we retain only one chiral half of the fermions, say
$\chi_\a$ and $\chi_{\bj\a}$, from (\ref{fermkin}).

The supersymmetry transformations (\ref{susyvars})
determine the transformations of the reduced fields which we also need
to re-express in our new set of variables.
From the variation (\ref{vsusy}) in $D=11$ we obtain
\be
{E_{(a}}^m \d E_{mb)} = i\bar\e \G_{(a} \Psi_{b)} \quad , \qquad
\d \phi = 2 i \bar\e \G^{10} \Psi_{10}.
\ee
For the first relation a compensating $SO(10)$ rotation for
maintaining the triangular gauge of the zehnbein
$\hat{E}_{\hat{m}}{}^{\hat{a}}$ was necessary.
Together with the redefinitions from eq.~(\ref{nscos1}), (\ref{Psitilde})
and (\ref{SUSYparameter}) this yields
\be
{e_{(a}}^m \d e_{mb)} = i \bar\ve \G_{(a} \tilde\Psi_{b)} \quad , \qquad
\d \vp = - i\bar\ve \G_{10} \tilde\Psi_{10}.
\ee
For the Lagrange multiplier $n= N\hat{G}^{-1/2} $, an analogous
calculation leads to the simple result
\be
\d n=-i\ve_\a\psi_{t\a}.
\ee
For the variation of the three form field $A_{mn\tten}$, we similarly obtain
\be
\d A_{mn\tten}  = \half \d b_{mn}
               = \half i\bar\e\G_{mn} \Psi_{\tten}
                 + i\bar\e \G_{\tten} \G_{[m} \Psi_{n]}
               =  i \bar\ve \G_{\tten} \G_{[m} \tilde\Psi_{n]}
\ee
Because the matrix $\G_{10}$ separates the two $SO(9)$'s (cf. appendix
\ref{gconv}.2), we can cast these variations into an $\soso$ covariant 
form by means of the redefined fermions and supersymmetry parameters 
(of course, retaining only one chiral half of (\ref{IIAfermions}))
and obtain
\begin{subequations}\label{Libj}
\be
  && {e_{(i}}^m \d e_{mj)} - \half{e_i}^m {e_j}^n \, \d
     b_{mn}
  = i \ve_\a\g_{i\a\b}\chi_{\bj\b} \equiv \Lambda_{i\bj}, \\
  &&   \d \vp = i\e_\a\chi_\a.
\ee
\end{subequations}
The variations of the metric $g_{mn}$ and the NS 2-form can thus be
recast into a manifestly $\soso$ covariant form
\be
\label{cosvar}
\d {\cal E}{\cal E}^{-1} = \Lambda_{i\bj}Y^{i\bj}.
\ee
whence
\begin{subequations}\label{varcosetfield}
\be
\d P_{i\bj} &=& D_t \Lambda_{i\bj} \equiv \pt
\Lambda_{i\bj}+Q_{ik}\Lambda_{k\bj}+Q_{\bj\bk}\Lambda_{i\bk}, \\
\d Q_{ij} &=& -2 \Lambda_{[i|\bk|}P_{j]\bk} \quad , \qquad
\d Q_{\bi\bj} = -2 \Lambda_{k[\bi}P_{|k|\bj]}.
\ee
\end{subequations}
where $D_t$ is a $\soso$ covariant derivative.

The variations of the fermions can be likewise determined from
eq.~(\ref{fsusy}), making use of all the redefinitions introduced above.
Again, one finds (after some computation) that all formulae can be cast
into a manifestly $\soso$ covariant form. Because the calculation
is completely analogous to the one for the bosonic fields, we refrain
from presenting further details here, but simply collect the pertinent
formulae in (\ref{SUSYfermions}) below.

{\em In summa}, we obtain the following Lagrangian for type I
reduced to one dimension with local $\soso$ invariance
\be
\cL_{{\rm I}}&=&\frac{1}{4}n^{-1}\big( P_{i\bj}P_{i\bj} -\dot\vp^2 \big)
-\frac{i}{2}\chi_\a D_t\chi_\a +\frac{i}{2}\chi_{\bj\a}D_t
\chi_{\bj\a}\nn\\
&&+\frac{i}{2} n^{-1}\psi_{t\a}\chi_\a{\dot\vp}
-\frac{i}{2} n^{-1} \psi_{t\a}\g_{i\a\b}\chi_{\bj\b}P_{i\bj}+\ldots
\ee
modulo higher order fermionic corrections. The derivatives on the
fermions are $\soso$ covariant, for instance
\be\label{sosocov}
D_t\chi_{\bi\a}=\partial_t\chi_{\bi\a}+
\ft{1}{4}Q_{jk}\g^{jk}_{\a\b}\chi_{\bi\b}
+Q_{\bi\bj}\chi_{\bj\a}.
\ee
The supersymmetry variations, which leave $\cL_{{\rm I}}$ invariant,
are, for the bosonic fields, \footnote{To make the formulae less
cumbersome, we will from now on suppress the $\soso$ spinor indices
whenever it is clear from the context which is meant.}
\begin{subequations}
\be
\d P_{i\bj} &=& D_t(i\ve\g_{i}\chi_{\bj}),\\
\d \vp &=& i\ve\chi, \\
\d n &=& -i\ve\psi_{t}.
\ee
\end{subequations}
The fermionic variations, with the cubic corrections needed
for the closure of the supersymmetry algebra below, read
\begin{subequations}\label{SUSYfermions}
\be
\d \psi_{t\a} &=& D_t \ve_\a
  - \frac{i}4 n (\g_{ij}\ve)_\a \, \chi\g_{ij} \chi +
   \frac{i}4 n (\g_{ij}\ve)_\a\, \chi_{\bk} \g_{ij} \chi_{\bk},\\
\d \chi_\a &=& -\frac{1}{2}n^{-1} \ve_\a (\dot\vp - i \psi_{t}\chi),\\
\d \chi_{\bj\a} &=& -\frac{1}{2}n^{-1}(\g_{i}\ve)_\a
   (P_{i\bj} - i\psi_{t}\g_{i}\chi_{\bj}).
\ee
\end{subequations}

Modulo the fermionic equations of motion, the supersymmetry algebra
closes and is $\soso$ covariant on all fields
\begin{align}
\label{susyI}
[\d_1,\d_2]n &=  \partial_t (\xi^t n), &
   [\d_1,\d_2]\psi_{t\a} &= 0,\nn\\
{}[\d_1,\d_2] \varphi &= \xi^t \partial_t \vp
    + \d_{\ve'} \vp, &
    [\delta_1,\delta_2] \chi_\a &= 0,\\
{}[\d_1,\d_2]P_{i\bj} &=  \partial_t (\xi^t P_{i\bj})
    +  \d_{\ve'} P_{i\bj} + \d_\omega P_{i\bj},&
    [\delta_1,\delta_2] \chi_{\bj\a} &= 0,\nn
\end{align}
where $\xi^t := -in^{-1} \ve_{2\a} \ve_{1\a}$ is the standard
(time) translation parameter, $\ve'_\a=-\xi^t\psi_{t\a}$ a new
local supersymmetry transformation parameter, and $\omega$
an $\soso$ transformation with parameters
\begin{subequations}
\be
\omega_{ij} &=&
\xi^t Q_{ij} - 2\chi_\bk \g_{i} \ve_{[1} \, \ve_{2]} \g_{j} \chi_\bk, \\
\omega_{\bi\bj} &=&
\xi^t Q_{\bi\bj} - 2\chi_k \g_{\bi} \ve_{[1} \, \ve_{2]} \g_{\bj]}
  \chi_k. 
\ee
\end{subequations}
This is precisely the form of the standard supersymmetry algebra
in theories of supergravity (in an on-shell formulation such as
ours): the commutator of two supersymmetry transformations yields
a translation term with parameter $\xi^\mu =i\ve_1 \g^\mu \ve_2$,
a local supersymmetry transformation with parameter
$\ve' = -\xi^\mu \psi_\mu$ (see e.g. \cite{vanN81}, section 1.9)
and a local gauge transformation, here in the form of a local $\soso$
transformation with parameters $(\omega_{ij},\omega_{\bi\bj})$.
The on-shell vanishing of the algebra (\ref{susyI}) on the
fermion fields is a consequence of this general structure in the
reduction to one dimension. For example, on the Lagrange
multiplier $\psi_t$, the contributions cancel by virtue of
\be
D_t \e'_\a + D_t (\xi^t \psi_{t\a}) = 0.
\ee
The closure on the matter fermions $\chi$ and $\chi_\bj$ requires
in addition the fermionic field equations
\begin{subequations}
\be
D_t \chi_\a + \frac12 n^{-1} \psi_{t\a} \dot\vp &=& 0, \\
D_t \chi_{\bj\a} + \frac12 (\g_i \psi_t)_\a P_{i\bj} &=& 0.
\ee
\end{subequations}

\end{subsection}

\begin{subsection}{\bf Type IIA theory}
\label{typeIIsec}

To extend type I to type IIA supergravity, we must complement the
type I fermions by those of the opposite chirality, and incorporate
the Kaluza--Klein vector $A_m$ together with the remaining components
of the 3-form field in eleven dimensions, namely $A_{mnp}$. In string
theory both arise from the Ramond-Ramond (RR) sector; for this reason
we will refer to them as RR type fields. With the Coulomb (temporal) 
gauge $A_t = A_{tmn} =0$, the time derivatives of the RR fields coincide 
with the field strengths of the unreduced theory
\be
F_{tmnp} \equiv \pt A_{mnp} = \half \pt a_{mnp} \quad ; \qquad
F_{tm} \equiv \pt a_m.
\ee
Repeating the calculation leading to (\ref{FF1}), the combined contribution 
of the Kaluza--Klein vector $a_m\equiv A_m$ from (\ref{redeinstein}) and
of $a_{mnp}\equiv 2A_{mnp}$ is thus equal to
\be
\label{typeIIred}
\frac{1}{4}n^{-1}e^{\vp} g^{\frac12}
              \left(\frac{1}{2} \cdot\frac{1}{3!}g^{mq} g^{nr} g^{ps}
               \pt a_{mnp} \pt a_{qrs} +\half
              g^{mn} \pt a_{m} \pt a_{n} \right),
\ee
Observe that the dilaton prefactor in (\ref{typeIIred}) is the
{\em same} for the 3-form and the Kaluza Klein vector, and is
indeed the desired prefactor for level $\ell =1$.
In addition, there is a common factor of $\sg$, which will turn out
to be precisely what is required for the $\sigma$-model (and can be
viewed as originating from a `spinorial metric' acting on the odd
level fields). As anticipated, our dilaton $\vp$ (defined by
Eq.~(\ref{nscos1})) couples in a way different from the standard
type IIA dilaton: in string frame, the latter does {\em not} appear
in front of the RR kinetic terms \cite{Po98}.

The RR form potentials thus give rise to the $SO(9)$ tensors $a^{(p)}$
for $p=1$ and $p=3$. By themselves, these fields are not enough
to allow for the larger symmetry $\soso$ or global $SO(9,9)$. However,
we can enhance the symmetry group in the desired way by adding forms
of degree $p=5,7$ and $9$. Namely, these fields can be then combined
into a {\em single} irreducible spinor representation of $SO(9,9)$,
which under the $\soso$ subgroup becomes the  $({\bf 16},{\bf 16})$
bispinor representation, and under the diagonal $SO(9)_\diag$ subgroup 
decomposes as
\be
\bf{16} \otimes \bf{16} = \bf{1} \oplus \bf{9} \oplus \bf{36} \oplus
   \bf{84} \oplus{126}
\ee
At the linearized level, we thus assemble the RR degrees of freedom into 
\be\label{phiaa}
\phi_{\a\ba}\equiv \phi_{\a\ba}^{{\rm (IIA)}}
=\sum_{p=1,3,5,7,9} \frac{1}{p!}\, \g^{a_1\ldots
a_p}_{\a\ba} a_{a_1\ldots a_p}^{(p)}.
\ee
As we will see, this is precisely the structure arising at level
$\ell =1$ in the decomposition of $E_{10}$ under its $SO(9,9)$
subgroup. Under the diagonal $SO(9)$ subgroup of $\soso$ we thus
recover the required representations for forms with
odd $p$. The 5- and 7-forms will be interpreted as the dual RR forms,
respectively, which in the reduced theory are associated to the
first order spatial gradients of $a_m$ and $a_{mnp}$. The 9-form,
on the other hand, does not appear in $D=11$ supergravity, but is
associated with the Romans' type mass term in the IIA theory
\cite{Ro86,BedeRoGrPaTo96,ObPi99}.
In contrast to the $A_9$ decomposition
of \cite{DaHeNi02}, where the fields and their duals appeared at
different levels, the level $\ell=1$ sector thus contains both
the RR type fields {\em and} their duals. By contrast, the dual
degrees of freedom for the $\ell =0$ NSNS fields appear only at
level $\ell =2$.
The formula (\ref{phiaa}) also admits a type IIB interpretation:
by means of the formulae of Appendix A.1 we can rewrite it as a sum
over {\em even} $p$,
\be\label{phiab}
\phi_{\a\ba}^{{\rm (IIB)}}
=\sum_{p=0,2,4,6,8} \frac{1}{p!}\, \g^{a_1\ldots
a_p}_{\a\ba} {\tilde a}_{a_1\ldots a_p}^{(p)}.
\ee
This rewriting simply reflects the equivalence of the type
IIA and IIB supergravity theories upon dimensional
reduction.\footnote{We note that our set of potentials is indeed
consistent with a type {\rm IIA} string theory interpretation
as RR potentials: The potentials of odd degree support Dirichlet
$p$-branes with $p$ even, and so $E_{10}$ `supports' $D0$, $D2$,
$D4$, $D6$, and $D8$ branes. In the type {\rm IIB} theory these are
transmuted into $D(-1)$, $D1$, $D3$, $D5$, and $D7$ branes, but
$E_{10}$ does not seem to support the space-filling $D9$-brane of
type {\rm IIB} string theory. However, there is also no known
ten dimensional massive {\rm IIB} supergravity theory containing 
the corresponding $10$-form (or scalar).}

In order to properly identify the RR degrees of freedom at the
non-linear level, we adopt a procedure that differs from \cite{DaHeNi02}
in so far as we deduce the relevant expressions from the supersymmetry
variations, and not by direct comparison with the (bosonic) equations
of motion. Instead, the latter, which we will study in section~4,
will provide us with an independent consistency check. Accordingly,
we proceed from the following ansatz for the fermionic variations
of the type II theory, encompassing the NSNS and RR degrees of freedom 
up to level $\ell =1$,
\begin{subequations}\label{SUSYII}
\be
\d \psi_{t\a} &=& D_t\ve_\a + e^{\frac12\vp}
  P_{\a\ba}\ve_\ba,\\
\d \psi_{t\ba} &=& D_t\ve_\ba - e^{\frac12\vp}
  \ve_\a P_{\a\ba},\\
\d\chi_\a &=& -\frac{1}{2}n^{-1}\dot\vp\ve_\a - \frac{1}{2}n^{-1}
    e^{\frac{1}{2}\vp} P_{\a\ba}\ve_\ba ,\\
\d\chi_\ba&=& -\frac{1}{2}n^{-1}\dot\vp\ve_\ba + \frac{1}{2}n^{-1}
    e^{\frac{1}{2}\vp} P_{\a\ba}\ve_\a ,\\
\d\chi_{i\ba} &=& -\frac{1}{2}n^{-1}P_{i\bj}\g_{\bj\ba\bb}\ve_\bb
    - \frac{1}{2}n^{-1} \ve_\b \g_{i\b\a} e^{\frac{1}{2}\vp}
    P_{\a\ba},\\
\d\chi_{\bj\a} &=& -\frac{1}{2}n^{-1}P_{i\bj}\g_{i\a\b}\ve_\b
    + \frac{1}{2}n^{-1} \ve_\bb \g_{\bj\bb\ba} e^{\frac{1}{2}\vp}
    P_{\a\ba},
\ee
\end{subequations}
where $P_{\a\ba}$ is the full nonlinear extension of the time 
derivative of the RR field $\phi_{\a\ba}$ in (\ref{phiaa}). (Below, 
we will separately discuss the level $\ell=2$ contributions to these 
variations, which correspond to the spatial gradients of the NSNS fields.) 
The terms involving NSNS degrees of freedom in (\ref{SUSYII}) are simply 
obtained by `doubling' the corresponding variations of the type~I
theory derived in the previous section. As for the RR type fields,
consistency with local $\soso$ invariance requires that they all 
appear via the {\em single} field strength $P_{\a\ba}$. For instance, 
the contributions from $F_{tmnp}$ to the variations of the redefined 
fermions are found to be (after some algebra, and using flat indices)
\begin{subequations}
\be
\d \tP_t & \ni & \frac{1}{12} \G^{abc}\ve \, F_{tabc}, \\
\d \tP_a &\ni& - \frac{1}{24} n^{-1} \G^{bcd}{\G_a} \ve \, F_{tbcd},\\
\d \tP_{10} &\ni& \frac{1}{24}n^{-1} \G_{10}\G^0 \G^{bcd}\ve\,
  F_{tbcd}.
\ee
\end{subequations}
Rewriting this in terms of curved indices and the redefined neunbein
and dilaton we obtain
\be
\G^{abc}\ve \, F_{tabc} &=& \G^{abc}\ve \, \big( {E_a}^m {E_b}^n {E_c}^p 
 F_{tmnp} + 3 {E_a}^m {E_b}^n {E_c}^\tten F_{tmn\tten} \big) \nn\\
&=& \G^{abc}\ve \, e^{\frac34\phi} e_a{}^m e_b{}^n e_c{}^p G_{tmnp} 
= \G^{abc} \ve \, e^{\frac12\vp} g^{\frac14} e_a{}^m e_b{}^n e_c{}^p G_{tmnp} 
\ee
with 
\be\label{Gtmnp}
G_{tmnp} := F_{tmnp} - 3 A_{[m} F_{np]t\tten}.
\ee
Comparing with (\ref{SUSYII}) we can read off the corresponding
contribution to $P_{\a\ba}$ and check that the field strength $G_{tmnp}$
(and hence also $F_{tmnp}$) indeed appears with {\em the same} coefficient 
in all variations.

The redefinition (\ref{Gtmnp}) is due to the reconversion from flat
back to curved indices by means of the rescaled neunbeine ${e_m}^a$ 
from (\ref{nscos1}). Together with the analogous redefinition
\be\label{Gmnpq}
G_{mnpq} := F_{mnpq} + 4 A_{[m} F_{npq]\tten},
\ee
it is well known from Kaluza Klein theory and ensures that these
field strengths do not transform under reparametrizations of the 
10-th spatial coordinate $\d x^{\tten} = \xi^{\tten}(t)$. Note that
the NSNS field strength $F_{tmn\tten}$ and the Kaluza--Klein fields
$F_{tm}$ and $F_{mn}$ do not receive any corrections of this form. 
The new
field strengths obey the modified Bianchi identities (cf. (\ref{Bianchi}))
\be\label{Bianchi1}
\pt G_{mnpq} - 4 \partial_{[m} G_{|t|npq]} &=& 4 F_{t[m} F_{npq]\tten}
 + 6 F_{[mn} F_{pq]t\tten}\nn\\[2mm]
&&\qquad + 4 A_{[m} \pt F_{npq]\tten}-12 A_{[m} \partial_n F_{pq]t\tten}
\ee

The remaining contributions from $F_{tm}$ and the (gauge invariant)
spatial gradients $F_{mn}$ and $G_{mnpq}$ are worked out similarly.
The Kaluza--Klein vector $A_m$ appears via the coefficients of 
anholonomity 
\begin{subequations}
\be
\Omega_{0a\, 10} &=& N^{-1} e^{\frac34 \phi} \cdot {e_a}^m F_{tm} \\
\Omega_{ab \, 10} &= & e^\phi \cdot {e_a}^m {e_b}^n F_{mn}
\ee
\end{subequations}
Some further calculation then yields the final result
\be\label{paa1}
P_{\a\ba} &=& g^{\frac14} \Bigg[ n e^{-\vp}M \d_{\a\ba} + 
\frac14 \g^m_{\a\ba} F_{tm}
   + \frac18 \g^{mn}_{\a\ba} n e^{-\vp} F_{mn} + \nn\\
&&  \qquad\qquad + \frac1{12} \g^{mnp}_{\a\ba} G_{tmnp} -
\frac1{48} \g^{mnpq}_{\a\ba} n e^{-\vp} G_{mnpq} \Bigg],
\ee
where we now and henceforth set $\g^m = \g^a e_a{}^m$ etc., using the 
redefined neunbein fields (\ref{nscos1}). The (constant) $M$ corresponds 
to a Romans type mass term, which exists only for type IIA supergravity 
in ten dimensions, but vanishes for $D=11$ supergravity. Note also the 
prefactors of $n e^{-\vp}$ in front of the terms containing even degree 
$\g$-matrices, {\sl i.e.} spatial gradients.

Although (\ref{paa1}) is the most convenient form to check against the
equations of motion and Bianchi identities (see section~4), it is also
straightforward to rewrite this expression in the form (\ref{phiaa}),
dualizing the 4-, 2- and 0-forms into 5-, 7- and 9-forms by means
of the formulae of the appendix. For instance, for the 4-form $G_{mnpq}$
we obtain (keeping track of the extra vielbein factor ${E_\tten}^{10}$!)
\be
G_{m_1\ldots m_4} &=& E_{m_1}{}^{a_1}\cdots E_{m_4}{}^{a_4}
  \e_{a_1\ldots a_4}{}^{b_1\ldots b_5} N^{-1}
  E_{b_1}{}^{n_1}\cdots E_{b_5}{}^{n_5} e^{-\frac12 \phi} {\tilde
  F}_{tn_1\ldots n_5 \tten}\nn\\ &=&
  N^{-1} e^{-\frac14 \phi} e_{m_1}{}^{a_1}\cdots e_{m_4}{}^{a_4}
  \e_{a_1\ldots a_4}{}^{b_1\ldots b_5} e_{b_1}{}^{n_1}\cdots e_{b_5}{}^{n_5}
  {\tilde F}_{tn_1\ldots n_5 \tten}
\ee
or
\be
G_{m_1 \ldots m_4} =
n^{-1} e^\vp \cdot {e_{m_1}}^{a_1} \cdots {e_{m_4}}^{a_4}
\e_{a_1\ldots a_4}{}^{b_1\ldots b_5}
 e_{b_1}{}^{n_1}\cdots e_{b_5}{}^{n_5} {\tilde F}_{tn_1\ldots n_5
  \tten}.
\ee
The prefactor $n^{-1} e^\vp$ here conveniently cancels the dilaton dependence
in front of $F_{mnpq}$ in (\ref{paa1}). Similarly, when dualizing
the exact Kaluza--Klein vector field strength by \footnote{The extra 
dilaton factor here is the one required by the equation of motion
for the Kaluza Klein vector $A_m$, as it follows from variation of 
the Einstein action in (\ref{maxsugra}) with (\ref{spatialvsplit}).}
\be
e^\phi F_{mn} = E_m{}^a E_n{}^b \e_{ab}{}^{c_1\ldots c_7}
  N^{-1}e^{-\frac12\phi} E_{c_1}{}^{m_1}\cdots E_{c_7}{}^{m_7}
  {\tilde F}_{tm_1\ldots m_7\tten},
\ee
we find
\be
F_{mn} &=& N^{-1} e^{-\frac14\phi} e_m{}^a e_n{}^b\e_{ab}{}^{c_1\ldots
  c_7} e_{c_1}{}^{m_1}\cdots e_{c_7}{}^{m_7}
  {\tilde F}_{tm_1\ldots m_7\tten}\nn\\
&=& n^{-1} e^\vp \cdot e_m{}^a e_n{}^b\e_{ab}{}^{c_1\ldots
  c_7} e_{c_1}{}^{m_1}\cdots e_{c_7}{}^{m_7}
  {\tilde F}_{tm_1\ldots m_7\tten}
\ee
again cancelling the prefactor in (\ref{paa1}). An analogous
calculation works for the prefactor of the Romans mass $M$ after
dualization to a $9$-form. Hence, we can rewrite the expansion
(\ref{paa1}) in the form
\be\label{paa2}
P_{\a\ba} &=&  g^{\frac14}\Bigg[ \frac14 \g^{m_1}_{\a\ba}F_{tm_1}
  +\frac1{12}\g^{m_1\ldots m_3}_{\a\ba} G_{tm_1\ldots m_3}
  -\frac1{48}\g^{m_1\ldots m_5}_{\a\ba} F_{tm_1\ldots
  m_5}\nn\\&&\qquad\qquad
  +\frac1{4\cdot 7!}\g^{m_1\ldots m_7}_{\a\ba} F_{tm_1\ldots m_7}
  +\frac1{9!}\g^{m_1\ldots m_9}_{\a\ba} F_{tm_1\ldots
  m_9}\Bigg],
\ee
where we have dropped additional $\tten$ indices and tildes on the dual
fields. This formula entails kinetic terms for the dual field strengths 
of the form
\be
n^{-1}e^\vp g^{m_1n_1}\cdots g^{m_pn_p} {F}_{tm_1\ldots m_p}
F_{tn_1\ldots n_p},
\ee
for $p=5,7,9$ supplementing and generalizing the terms in (\ref{typeIIred}).

After these preparations we are now ready to give the supersymmetry
variations of the bosonic degrees of freedom, as well as the type II
Lagrangian. The variation of the bosonic type I fields are again
obtained by `doubling' since we now have to include two sets
of fermions, and read
\begin{subequations}
\be
\d n&=&-i(\ve_\a\psi_{t\a}+\ve_\ba\psi_{t\ba}),\\
\d\vp&=&i(\ve_\a\chi_\a+\ve_\ba\chi_\ba),\\
\d P_{i\bj}&=& D_t(i\ve_\a\g_{i\a\b}\chi_{\bj\b}
    +i\ve_\ba\g_{\bj\ba\bb}\chi_{i\bb}).
\ee
\end{subequations}
From the variations (\ref{vsusy}) and (\ref{gsusy}) we furthermore
deduce the following transformation of the RR fields
\be\label{dPaa}
\d \big(e^{\frac{1}{2}\vp}P_{\a\ba}\big)
    &=&  2 D_t \Big[ i(\ve_\a\chi_\ba
    -\ve_\ba\chi_\a)+i(\ve_\b\g_{i\b\a}\chi_{i\ba}
    -\ve_\bb\g_{\bj\bb\ba}\chi_{\bj\a})\Big].
\ee
Note the minus signs and the dilaton coupling in front of the RR terms;
we will always treat $e^{\frac{1}{2}\vp}P_{\a\ba}$ as a field on the
same footing as the type I fields.

The $\soso$ invariant Lagrangian for the type II theory requires
keeping both chiralities of the reduced fermion kinetic term
(\ref{fermkin}) and appropriate extra Noether terms involving
fermions of opposite chirality and the new field strength
$e^{\frac12\vp}P_{\a\ba}$. Like the type I Lagrangian and the 
fermionic variations of the RR fields, the extra terms in this 
Lagrangian can be checked against the corresponding terms obtained 
by directly reducing (\ref{maxsugra}); here again, the redefinitions 
of the fields found before are essential in order to obtain complete 
agreement. Our construction thus makes use of both dimensional reduction
and the supersymmetric completion of the `doubled' type~I Lagrangian 
by means of the supersymmetry transformations (\ref{SUSYII}).
The result is (modulo higher fermionic terms)
\be\label{typeIIlag}
\cL_{{\rm II}} &=& \frac{1}{4}n^{-1}\left(
P_{i\bj}P_{i\bj} - {\dot\vp}^2\right)+\frac{1}{8}
n^{-1} e^{\vp} P_{\a\ba}P_{\a\ba}\nonumber\\
&&-\frac{i}{2}\chi_\a D_t\chi_\a +\frac{i}{2}\chi_{\bj\a}D_t
    \chi_{\bj\a}-\frac{i}{2}\chi_\ba D_t\chi_\ba
    +\frac{i}{2}\chi_{i\ba}D_t \chi_{i\ba}\nn\\
&& -i e^{\frac12 \vp} \chi_\a\chi_\ba P_{\a\ba} +
  i e^{\frac12 \vp} (\chi_\bj \g_i)_\a (\g_\bj \chi_i)_\ba P_{\a\ba}\nn\\
&&+\frac{i}{2} n^{-1}\big(\psi_{t\a}\chi_\a+\psi_{t\ba}\chi_\ba\big){\dot\vp}
    -\frac{i}{2} n^{-1}\big(\psi_{t\a}\g_{i\a\b}\chi_{\bj\b} P_{i\bj}
     + \psi_{t\ba}\g_{\bi\ba\bb}\chi_{j\bb} P_{\bi j})\big)\nonumber\\
&& +\frac{i}{2}n^{-1} \big( \psi_{t\ba} \chi_\a - \psi_{t\a} \chi_\ba
    +\psi_{t\bb} \g_{\bj  \bb\ba} \chi_{\bj \a}
    -\psi_{t\b} \g_{i  \b\a} \chi_{i\ba} \big)
    e^{\frac{1}{2}\vp}P_{\a\ba}+\ldots
\ee
which is manifestly invariant under local $\soso$
transformations. From (\ref{typeIIlag}), we can also check that
our identification (\ref{paa1}) for the decomposition of $P_{\a\ba}$
with all numerical coefficients is
consistent with the reduction of the kinetic terms for the three
form and the Kaluza--Klein vector given in (\ref{typeIIred}).

When studying the supersymmetry of the Lagrangian (\ref{typeIIred})
up to linear order in the fermions, we find that many cancellations
work as before, but there are also some new features. For instance,
variation of the new Noether term coupling fermions of opposite
chirality produces (amongst other terms) a contribution
\be
\frac{i}4 n^{-2} e^\vp P_{\a\ba} P_{\b\ba}
\Big[ - \psi_{t\a} \ve_\b + (\psi_t \g_j)_\a (\g_j \ve)_\b \Big]
\ee
and an analogous term with the opposite chiralities. After an
$SO(9)$ Fierz rearrangement this becomes
\be
\frac{i}8 n^{-2}  e^\vp P_{\a\ba} P_{\b\ba} \Big[ (\psi_t \ve)
\cdot \d_{\a\b} - (\psi_t \g_j \ve) \cdot \g^j_{\a\b} \Big]
\ee
The first term cancels against the variation of $n$ in the `kinetic
term' for $P_{\a\ba}$, but there is no other contribution to cancel
the second term. Instead this term vanishes by itself on account of
the specific form (\ref{paa2}) of $P_{\a\ba}$: because there are only
odd order $\g$-matrices, the contributions that survive in the trace
vanish by symmetry.

Secondly, we have no complete cancellation of all terms (even neglecting
higher order fermionic terms), and therefore no complete supersymmetry.
This is because we are left with the following terms
\be\label{typeIImismatch}
\d \cL_{{\rm II}} &=& -\frac{i}{2}n^{-1} \g^{ik}_{\a\b}\ve_\b
  P_{k\bj}\g_{\bj\ba\bb}\chi_{i\bb}e^{\frac12\vp}P_{\a\ba}
  -\frac{i}{2}n^{-1} \g^{\bj\bk}_{\ba\bb}\ve_\bb
  P_{i\bk}\g_{i\a\b}\chi_{\bj\b}e^{\frac12\vp}P_{\a\ba}\nn\\
&&+\frac{i}{2}n^{-1}e^\vp (\ve\g_\bj)_\ba P_{\a\ba}\g_{i\a\b}P_{\b\bg}
  (\g_\bj\chi_i)_\bg -\frac{i}{2}n^{-1}e^\vp (\g_i\chi_\bj)_\a
  P_{\a\ba}\g_{\bj\ba\bb}P_{\b\bb}(\g_i\ve)_\b\nn\\
&&-\frac{i}{2}n^{-1}e^\vp \ve_\bb
 P_{\b\bb}\g_{i\b\a}\chi_{i\ba}P_{\a\ba} -\frac{i}{2}n^{-1}e^\vp\ve_\b
 P_{\b\bb}\g_{\bj\bb\ba}\chi_{\bj\a}P_{\a\ba}.
\ee
These terms apparently cannot be cancelled with the present fields,
because the only coupling between $\chi_i$, $\chi_\bj$ and $P_{\a\ba}$
is the one given in the Lagrangian. We interpret this as an indication
of the need to introduce additional fermionic fields for the type II
theory, a fact which is further supported by an analysis of the
supersymmetry algebra.

On investigating the closure of the supersymmetry algebra we find that
it no longer closes into time translations and gauge transformations as
the type I superalgebra, even disregarding higher order fermionic
contributions. For instance, the commutator on the RR fields gives
\be
[\d_1 , \d_2] \big(e^{\frac{1}{2}\vp}P_{\a\ba}\big)
    &=& iD_t\Big[n^{-1} \Big( \ve_{2\a} \ve_{1\b}
    e^{\frac{1}{2}\vp}P_{\b\ba}
    + \ve_{2\ba} \ve_{1\bb}e^{\frac{1}{2}\vp} P_{\a\bb} \Big) \nn\\
&& - n^{-1} \big( \ve_{2\b} \g_{i\b\a} \ve_{1\d} \g_{i\d\g}
    e^{\frac{1}{2}\vp}P_{\g\ba}
    + \ve_{2\bb} \g_{\bj\bb\ba} \ve_{1\bd} \g_{\bj\bd\bg}
    e^{\frac{1}{2}\vp}P_{\a\bg}\big)\Big].
\ee
minus a term with $\ve_1$ and $\ve_2$ exchanged (and a term where
the gauge fields are varied).
After an $SO(9)$ Fierz rearrangement, we obtain
\be\label{rrsusyalg}
[\d_1 , \d_2] \big(e^{\frac{1}{2}\vp}P_{\a\ba}\big)
    &=& D_t\Big[ \xi^t e^{\frac{1}{2}\vp}P_{\a\ba}
    - \xi^i\g_{i\a\b}e^{\frac{1}{2}\vp}P_{\b\ba}
    - \xi^\bj\g_{\bj\ba\bb}e^{\frac{1}{2}\vp}P_{\a\bb}\Big].
\ee
Here
\be
\xi^t=-i n^{-1}(\ve_{2\a}\ve_{1\a}+\ve_{2\ba}\ve_{1\ba})
\ee
is the expected (doubled) time translation parameter. In addition, we
now have the two new parameters
\be
\xi_i=-i n^{-1} \ve_{2\a}\g_{i\a\b}\ve_{1\b} \quad ; \qquad
\xi_\bj=-i n^{-1} \ve_{2\ba}\g_{\bj\ba\bb}\ve_{1\bb}
\ee
which look like {\em spatial} translations! The correct interpretation 
of these terms remains an open problem for the time being.

We conclude this subsection with some comments on the level $\ell=2$ 
sector of the $E_{10}$ coset model which contains the spatial 
gradients of the NSNS fields, and which we have analysed only
partially (see also section \ref{sigmamod}). As we saw above, 
the supersymmetry  variations (\ref{SUSYII}) of the fermions only 
include the contributions from the NSNS fields, but not their dual 
fields, whereas the RR sector at $\ell =1$ contains {\em both} 
the time derivatives of the RR fields {\em and} their first order
spatial gradients via $P_{\a\ba}$. At level $\ell =2$, the relevant 
representation of $SO(9,9)$ is an antisymmetric $3$-tensor 
$P_{IJK}\equiv P^{(2)}_{IJK}$ containing the 
spatial components $\o_{a \, bc}$ of the spin connection and 
the (gauge invariant) spatial gradients $F_{mnp\tten}$ 
of the NSNS $2$-form $b_{mn}$. As with the $A_9$ decomposition
of \cite{DaHeNi02}, we cannot so far accommodate the trace
components $\o_{a \, ab}$ and $\o_{10 \, 10b}$ (the latter being
directly related to the spatial gradients $\partial_i \vp$ of the 
dilaton $\vp$). We will thus consider only the traceless part 
\be
\to_{a\,bc} = \o_{a\,bc}+
   \frac18 ( \d_{ab} \o_{d\,cd} - \d_{ac} \o_{d\, bd}).
\ee
which contains the fully antisymmetric part $\to_{[a \, bc]}$
as well as a mixed Young tableau representation.

As before, we determine the relevant expressions for the NSNS gradients
from the supersymmetry variations (\ref{fsusy}), rather than the 
equations of motion. 
Evaluating (\ref{fsusy}), we find the following contributions
\begin{subequations}
\label{ell2susy}
\be
\d \tP_{10} &\ni& \hG^{\frac12}\Big(-\frac14\to_{a\,bc}
\G_{10}\G^{abc} -\frac1{12} F_{10abc} \G^{abc}\Big)\ve\\
\d \tP_a &\ni& \hG^{\frac12}\Big(\frac14\to_{a\, bc}
\G^{bc} -\frac1{4} F_{10abc} \G^{10}\G^{bc}\Big)\ve,\\
\d \tP_t &\ni& N\Big(-\frac14\to_{a\,bc}
\G_{0}\G^{abc} -\frac1{12} F_{10abc} \G_0\G^{10}\G^{abc}\Big)\ve.
\ee
\end{subequations}
We note that, converting back to curved indices, the prefactors
are of the desired form; for instance, the coefficient of the
two-form gradient $F_{mnp\tten}$ above comes out to be
\be
\hG^{\frac12}e^{\frac14\phi} = e^{-\vp}.
\ee
We make the following ansatz for the $\ell=2$ contributions $\d^{(2)}$
to the supersymmetry variations:
\begin{subequations}
\label{lev2vars}
\be
\d^{(2)} \chi_\a  &=&  n^{-1}e^\vp P_{ijk} \g^{ijk}_{\a\b}\ve_\b,\\
\d^{(2)} \chi_\ba &=& - n^{-1}e^\vp P_{\bi\bj\bk}
  \g^{\bi\bj\bk}_{\ba\bb}\ve_\bb,\\
\d^{(2)} \chi_{\bi\a} &=&  n^{-1}e^\vp
   P_{\bi jk} \g^{jk}_{\a\b}\ve_\b,\\
\d^{(2)} \chi_{i\ba} &=&- n^{-1}e^\vp
   P_{i\bj\bk} \g^{\bj\bk}_{\ba\bb}\ve_\bb,\\
\d^{(2)} \psi_{t\a} &=& e^\vp P_{ijk} \g^{ijk}_{\a\b}\ve_\b,\\
\d^{(2)} \psi_{t\ba} &=& -e^\vp P_{\bi\bj\bk}
  \g^{\bi\bj\bk}_{\ba\bb}\ve_\bb.
\ee
\end{subequations}
Note that these are indeed the only $\soso$-covariant expressions 
one can write down for the variations.
Combining the ansatz with our results (\ref{ell2susy}), we can read
off the expressions for the
components of $P_{IJK}$ in the $\soso$ decomposition, {\em viz.}
\begin{subequations}
\be
P_{ijk} &=& n e^{-2\vp} e_i{}^m e_j{}^n e_k{}^p
\left( \frac14 \to_{m \, np} - \frac1{12} F_{mnp\tten} \right),\\
P_{ij\bk} &=& n e^{-2\vp} e_i{}^m e_j{}^n e_k{}^p \left(
\frac14 \o_{m\, np} + \frac1{4} F_{mnp\tten} \right),\\
P_{i\bj\bk} &=& n e^{-2\vp} e_i{}^m e_j{}^n e_k{}^p
\left( \frac14 \to_{m \, np} - \frac14 F_{mnp\tten} \right) ,\\
P_{\bi\bj\bk} &=& n e^{-2\vp} e_i{}^m e_j{}^n e_k{}^p 
\left( \frac14 \to_{m \, np} + \frac1{12} F_{mnp\tten} \right).
\ee
\end{subequations}
(Only the components with both barred and unbarred indices contain 
the mixed Young tableau representation in $\o_{m\, np}$.)

We will check these results against the bosonic equations of motion 
in section \ref{sigmamod}. Observe also that including the NSNS 
gradients might affect the closure of the supersymmetry algebra, 
but cannot affect the extra terms (\ref{typeIImismatch}) in the 
variation of the Lagrangian.

Finally, the appearance of the combination
\be
\d \psi_{t\a} = D_t \e_\a + e^{\frac12\vp}P_{\a\ba}\e_\ba + e^\vp
P_{ijk}\g^{ijk}_{\a\b}\e_\b.
\ee
in the supersymmetry variations (\ref{SUSYII}) combined with
(\ref{lev2vars}) 
signals the beginning of an enlargement of the $\soso$ covariant
derivative to a derivative which is also covariant w.r.t. the 
RR fields $e^{\frac{1}{2}\vp}P_{\a\ba}$ on level 1 and the level 2 fields
$e^\vp P_{IJK}$. These should be interpreted as the first terms
beyond the $\soso$ covariantization in a $K(E_{10})$ covariant
derivative acting on the $K(E_{10})$ spinor representation of
which the fermions above make up a small part (the grading being kept
track of by the dilaton factors). Similar recombinations occur in the
Lagrangian (\ref{typeIIlag}) for the kinetic terms of the fermion fields.

\end{subsection}

\end{section}

\begin{section}{\bf Level decomposition of \bmath $\lae_{10}$
under its
$\mathfrak{so}(9,9)$ \ubmath subalgebra}
\label{e10rels}
\setcounter{equation}{0}

\begin{figure}
\begin{center}
\includegraphics{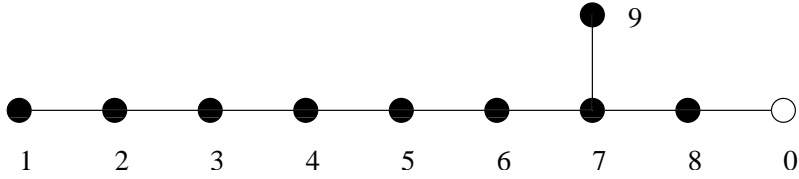}
\caption{\label{eddynk} The Dynkin diagram of $\lae_{10}$ with the
  regular $D_9$ subalgebra indicated by the solid nodes.}
\end{center}
\end{figure}

In this section we prepare the analysis of the $E_{10}/K(E_{10})$
$\sigma$-model by studying the decomposition of the hyperbolic
KM algebra $\lae_{10}$ under its $D_9\equiv \mathfrak{so}(9,9)$
subalgebra. This will pave the way for the next section, where
we will establish the matching between the equations of motion
of the $E_{10}/K(E_{10})$ at the first two levels with the
appropriately truncated bosonic equations of motion of $D=11$
supergravity. This matching is analogous to the one obtained w.r.t.
the $A_9$ decomposition in \cite{DaHeNi02}. The $D_9$ decomposition
of $E_{10}$ also allows for a simple proof that the second
simply-laced maximal rank hyperbolic Kac Moody algebra $DE_{10}$ is
actually a proper subalgebra of $E_{10}$ --- in agreement with
the embedding of type I into type II supergravity.

The (split) algebra $\lae_{10}$ is the Kac-Moody algebra defined by the Dynkin
diagram of figure \ref{eddynk}, where we have marked the nodes which
define the regular subalgebra of type $D_9$, numbering them from
$i=1,...,9$; the remaining node will be labeled by `0'. The $D_9$
subalgebra appears not in its compact form $\mathfrak{so}(18)$, but via its
split form $\mathfrak{so}(9,9)\subset\lae_{10}$ --- in fact,
the compact algebra $\mathfrak{so}(18)$ is {\em not} a
subalgebra of $\lae_{10}$ (whereas $\mathfrak{so}(16)\subset
\mathfrak{e}_{8(8)}$ obviously is). The algebra $\lae_{10}$ can now
be decomposed into an infinite sum of irreducible representations
of $\mathfrak{so}(9,9)$. For the determination of the latter one can
use the same techniques as the ones that were used in \cite{NiFi03}
to work out the level decomposition of $\lae_{10}$ w.r.t. its $A_9$
subalgebra to rather high levels. The results are tabulated in
appendix \ref{eddec}, which also gives the {\em outer multiplicities}
of the relevant representations, i.e. the number of times
they appear at a given level. Observe that at even levels
we have only vectorial representations, whereas at odd levels all $SO(9,9)$
representations are spinorial. Under the $\soso$ subgroup these appear
as the product of two spinorial $SO(9)$ representations and therefore
as tensorial (single-valued) representations of the diagonal $SO(9)$.
We will thus associate the even level representations with the NSNS type
fields, and the odd level ones with the RR type fields\footnote{For
the $A_9$ decomposition, there is no such distinction because $A_9$
does not admit (finite dimensional) spinor representations. Similarly,
$SO(9,9)$ does not admit (finite dimensional) representations which
decompose into spinors of the diagonal $SO(9)$ subgroup.}.

\begin{subsection}{\bf \bmath $E_{10}$ \ubmath at low levels}

We first spell out the relation between the $\lae_{10}$ Chevalley
generators and the $SO(9,9)$ generators used in the supersymmetric
$\s$-model description. To begin with, we would like to identify the
$D_9$ Chevalley generators $e_i, f_i$ and $h_i$ with $i=1, \ldots, 9$
in terms of the $SO(9,9)$ generators $M^{IJ}=-M^{JI}$ obeying the
standard commutation relations
\be
\left[ M^{IJ},M^{KL}\right]=
\eta^{KI}M^{JL}-\eta^{KJ}M^{IL} -\eta^{LI}M^{JK}+\eta^{LJ}M^{IK},
\ee
with
\be\label{eta}
\eta^{IJ} = {\rm diag}\,\big[ (+)^9,(-)^9\big] \quad\Longleftrightarrow\quad
\eta^{ij}=\d^{ij} = -\eta^{\bi\bj} \;\; ; \;\; \eta^{i \bj}=\eta^{\bi j}=0
\ee
where we made use of the $SO(9)\times SO(9)$ indices $I\equiv (i,\bi)$
already introduced before. We use $\eta$ to raise and lower indices in the
standard fashion. With the Cartan--Killing form
\be\label{CK3}
(M^{IJ}|M^{KL}) = \eta^{KJ}\eta^{IL}-\eta^{KI}\eta^{JL}
\ee
we can split the generators into compact and non-compact ones
\be
X^{ij} = M^{ij} \quad , \quad X^{\bi\bj} = -M^{\bi\bj}
\quad ; \qquad Y^{i\bj} = M^{i\bj}.
\ee
The inner product of the non-compact generators is then
\be
(Y^{i\bj}|Y^{k\bl})=\d^{ik}\d^{\bj\bl}.
\ee
The $\mathfrak{so} (9,9)$ Chevalley generators are obtained
by setting (recall that $\bi \equiv i + 8$, etc.)
\begin{subequations}
\be
e_i &=& \frac12 \Big(M^{\overline{i+1},\bi}- M^{i+1,i}
      +  M^{i+1,\bi}-  M^{\overline{i+1},i} \Big) \\
f_i &=& \frac12 \Big(M^{i+1,\bi}-M^{\overline{i+1},i}
           - M^{\overline{i+1},\bi} + M^{i+1,i} \Big)\\
h_i&=&M^{i+1,\overline{i+1}}-M^{i,\bi}
\ee
\end{subequations}
for $i=1, \ldots, 8$ and
\begin{subequations}
\be
e_9 &=& \frac12 \Big( M^{\overline{9},\overline{8}}+M^{9,8}
      -M^{9,\overline{8}} -  M^{\overline{9},8} \Big)  \\
f_9 &=& - \frac12 \Big( M^{9,\overline{8}} + M^{\overline{9},8}
           + M^{\overline{9},\overline{8}}+ M^{9,8}\Big) \\
h_9 &=&-M^{8,\overline{8}}-M^{9,\overline{9}}.
\ee
\end{subequations}
With the $SO(9,9)$ commutation relations given above it is
straightforward to check that they indeed satisfy the generating
relations. Furthermore, the standard invariant bilinear form is
precisely the one given by (\ref{CK3}).

The remaining independent generator at level $\ell =0$ is the
Cartan subalgebra generator $T$ corresponding to the fundamental 
weight $\lambda_{0}$ associated to the node marked $0$ in 
figure \ref{eddynk}, whose explicit form is
\be
T=-h_1-2h_2-3h_3-4h_4-5h_5-6h_6-7h_7-\frac{9}{2}h_8-\frac{7}{2}h_9-2h_0.
\ee
It generates a $GL(1)\equiv\reals^+$ subgroup which commutes
with and is orthogonal to the $\mathfrak{so}(9,9)$ subalgebra, and is
normalized according to
\be
\label{so99inv}
(T|M^{IJ})= 0,\quad , \qquad (T|T)= -1.
\ee

The maximal compact subalgebra is the invariant subalgebra w.r.t.
the Chevalley involution
\be
\o (e_i) = - f_i \;\; , \quad
\o (f_i) = - e_i \;\; , \quad
\o (h_i) = - h_i \;\; ; \quad
\ee
and
is therefore spanned by the multiple commutators of the $(e_i - f_i)$.
From
\begin{subequations}
\be
e_i-f_i &=& M^{\overline{i+1},\bi}-M^{i+1,i}\nn\\
e_9-f_9 &=& M^{\bar{9},\bar{8}}+M^{9,8}.
\ee
\end{subequations}
we see that the combinations $(e_8-f_8)\pm (e_9-f_9)$ are
Lorentz generators solely containing barred $M^{\bar{9},\bar{8}}$
or unbarred indices $M^{9,8}$, respectively. Under further commutation
with the remaining $(e_i-f_i)$ we can thus obtain any Lorentz generator
with only barred or unbarred indices. This identifies the maximal compact
subalgebra as $\mathfrak{so}(9)\oplus\mathfrak{so}(9)$, as anticipated.

Performing a decomposition of $\lae_{10}$ into representations of
$\mathfrak{so}(9,9)$ we obtain the table given in appendix \ref{eddec}
whose first four entries we reproduce here for convenience. (All fields
occur with outer multiplicity one.)
\begin{center}
\begin{tabular}{cccc}
$\ell$&$[p_1\ldots p_9]$&$SO(9,9)$ generator&Transposed generator at level
$-\ell$\\[2mm]
\hline\\
0&[010000000]&$M^{IJ}
$&$ M_{IJ} = \o(M^{IJ})
$\\
0&[000000000]&$T$&$T$\\
1&[000000010]&$E_A$&$F_\dA$\\
2&[001000000]&$E^{IJK}$&$F_{IJK}$\\[2mm]
\end{tabular}
\end{center}

Having already discussed $\ell=0$ above, we
see from the table, there is only one representation at
level $\ell =1$, with Dynkin label $[000000001]$. This is the
${\bf 256_s}$ spinor representation $E_A$ of $\mathfrak{so}(9,9)$; the
Chevalley generator $e_{0}$ of $\lae_{10}$ corresponds to the highest
weight state of this representation. The components of the spinor $E_A$
carry $GL(1)$ charge $\frac{1}{2}$ with our choice for $T$. Using the
Weyl spinor notation of appendix \ref{gconv}, we thus have
\begin{subequations}
\be
\left[M^{IJ},E_A\right]&=&-\ft{1}{2}\S^{IJ}_{AB} E_B,\\
\left[T,E_A\right]&=&\ft{1}{2} E_A.
\ee
\end{subequations}

The conjugate generators $-\o (E_A)=\CC_{A\dB} F_\dB$ at level $\ell=-1$
belong to the conjugate spinor representation
$\bf{256}_c$.\footnote{Here $\CC$ pertains to the charge conjugation
matrix, see appendix \ref{gconv} for notation.} This follows
from the fact that {\em the Chevalley involution reverses chirality},
which itself is a consequence of the fact that the chirality operator
$\wG^*$ (see appendix \ref{gconv}) is represented as the product over an odd
number of Cartan subalgebra generators in terms of $\wG^{IJ}$, which implies
\be
\o (h_1 \cdots h_9) = - h_1 \cdots h_9
\ee
which in turn requires that on the spinor representations
\be
\o (\wG^*) = - \wG^*
\ee
Using dotted indices, we have the commutation relations
\begin{subequations}
\be
\left[M^{IJ},F_{\dA}\right] &=& -\ft12 \bar{\S}^{IJ}_{\dA\dB} F_{\dB} , \\
\left[T,F_{\dA}\right]&=&-\ft{1}{2} F_{\dA}.
\ee
\end{subequations}
The commutator of level $\ell=1$ with level $\ell=-1$ is
\be
\left[E_A,F_{\dB}\right]&=&\frac{1}{4}(\S^{IJ}\CC)_{A{\dB}}M_{IJ}
-\frac{1}{2}\CC_{A{\dB}}T,
\ee
where the constants are determined from
the inner product of these fields within $\lae_{10}$
\be
(E_A|F_{\dB})=\CC_{A\dB}.
\ee

At $\ell=2$ we have the $\mathfrak{so}(9,9)$ representation with
Dynkin label $[001000000]$, {\sl i.e.} a 3-form generator $E^{IJK}$ with
commutation relations
\begin{subequations}
\be
\left[M^{IJ},E^{KLM}\right]&=&3\eta^{I[K}E^{LM]J}
                            - 3\eta^{J[K}E^{LM]I}, \\
\left[T,E^{KLM}\right]&=&E^{KLM}.
\ee
\end{subequations}
It is generated by taking the commutator of two $\ell=1$ generators
\be\label{comm11}
\left[E_A,E_B\right]&=&\frac16 ({\S}_{IJK}\CCb)_{AB}E^{IJK}.
\ee
Note that $\S^{IJK}\CCb$ is indeed antisymmetric
as required for consistency. The transposed field
\be
F_{IJK}:=-\omega(E^{IJK})
\ee
(note the position of indices) satisfies
\begin{subequations}
\be
\left[M^{IJ},F^{KLM}\right]&=&3\eta^{I[K}F^{LM]J} -
                           3\eta^{J[K}F^{LM]I}  \\
\left[T,F^{KLM}\right]&=& -F^{KLM}\\\label{comm-1-1}
\left[F_{\dA},F_{\dB}\right]&=&-\frac16 (\bar{\S}^{IJK}\CC)_{\dA\dB}F_{IJK}
\ee
\end{subequations}
To determine the proper normalization of the level 2 generators,
we observe that the combination
\be
v = E^{123}+
E^{\bar{1}23}+E^{1\bar{2}3}+E^{12\bar{3}}
+E^{1\bar{2}\bar{3}}+E^{\bar{1}2\bar{3}}+E^{\bar{1}\bar{2}3}
+E^{\bar{1}\bar{2}\bar{3}} \; ,
\ee
is the lowest weight vector of the representation with basis 
elements $E^{IJK}$. Normalizing it according to $-(v|\o (v))=1$
we deduce the inner product of $E^{IJK}$ and $F_{LMN}$ 
\be
(E^{IJK}|F_{LMN})&=&\frac34 \, \delta^{IJK}_{LMN},
\ee
This normalization was also used to fix the constants in (\ref{comm11}) 
and (\ref{comm-1-1}).

The remaining commutators up to level $\ell=2$ are
\begin{subequations}
\be
\left[E^{IJK},F_\dA\right] &=&
    \frac{1}{8}\S^{IJK}_{B\dA} E_B,\\
\left[F^{IJK},E_A\right] &=&
    -\frac{1}{8}\bar{\S}^{IJK}_{\dB A} F_\dB,\\
\left[E^{IJK},F_{LMN}\right] &=&-\frac34
    \,\d^{IJK}_{LMN} T -\frac94 \d^{[IJ}_{[LM} {M^{K]}}_{N]}.
\ee
\end{subequations}

Finally, let us have a look at the `affine  representations' analogous
to those identified in \cite{DaHeNi02}, and proposed there to be associated
to the spatial gradients. In the present decomposition, they are ($n\geq 0$)
\begin{subequations}
\be
\ell = 2n+1 \qquad &\longleftrightarrow& \qquad [n00000010] \\
\ell = 2n+2 \qquad &\longleftrightarrow& \qquad [n01000000]
\ee
\end{subequations}
Like the corresponding representations in \cite{DaHeNi02}, they
appear all with outer multiplicity one. Trying a similar
interpretation, we are led to associate ($I=1,\ldots, 18$)
\begin{subequations}
\be
\p_{I_1}\cdots \p_{I_n} P_{\a\ba} \qquad &\longleftrightarrow&
\qquad [n00000010], \\
\p_{I_1}\cdots \p_{I_n} P_{JKL} \qquad &\longleftrightarrow& \qquad
[n01000000],
\ee
\end{subequations}
since the generator $E_A-\CC_{A\dA}F_\dA$ multiplies a field
strength $P^A= P_{\a\ba}$ under $\soso$.
As these are representations of $SO(9,9)$ they come with an
additional tracelessness constraint, for example
\be
\hat{\G}^I_{\a\ba \, \b\bb}  \p_I P_{\b\bb} = 0.
\ee
Notice that these are now 18-dimensional `gradients', and therefore
an interpretation along the lines of \cite{DaHeNi02} is more subtle.
Splitting the $18$ components into $9+9$, and recalling that the two
$SO(9)$ groups act on left and right moving sectors of the superstring,
respectively, we are led to tentatively associate these representations
to the derivatives w.r.t. the left and right moving target 
space coordinates $x^i$ and $x^\bi = \tilde x^i$ (where 
the latter are defined by the worldsheet duality relation 
$\p_\pm x^i = \pm \p_\pm \tilde x^i$). We see no direct 
trace of the derivatives w.r.t. the circle in the eleventh 
direction, but we know from the $GL(10)$ analysis of
\cite{DaHeNi02} that they are present.

\end{subsection}
\begin{subsection}{\bf \bmath Type I $\subset$ Type II, and
$DE_{10}\subset E_{10}$
\ubmath}

Type I supergravity is a subsector of the type II theory and is
conjectured to possess a hidden Kac--Moody symmetry $DE_{10}$ in
one dimension \cite{Ju82}. The Dynkin diagram of the hyperbolic
over-extension $DE_{10}=D_8^{++}$ is displayed in figure
\ref{deddynk}. Here we show that the conjectured $DE_{10}$
symmetry is completely consistent with the $E_{10}$ symmetry of
the type IIA theory by proving that $\mathfrak{de}_{10}$ is a subalgebra of
$\lae_{10}$ \footnote{This shows that there is really {\em only one}
simply laced maximal rank 10 hyperbolic Kac Moody algebra, a fact
which seems to have gone unnoticed in the literature!}. More
specifically, the elements $DE_{10}$ form a subset of the NSNS
sector of $E_{10}$, corresponding to the even levels in the $D_9$
decomposition of $E_{10}$. We note that the analogous embedding of
$DE_{11}$ into $E_{11}$ was recently established in
\cite{SchnWe04} also by invoking the embedding of type I into type
IIA supergravity, but the level decomposition via their common
$D_{10}$ has so far not been studied in any detail.

\begin{figure}
\begin{center}
\includegraphics{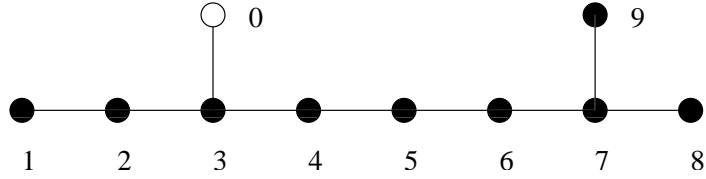}
\caption{\label{deddynk} The Dynkin diagram of $DE_{10}=D_8^{++}$ with the
  regular $D_9$ subalgebra indicated by the solid nodes. This subalgebra
  is identical to the one of $E_{10}$ in figure \ref{eddynk}.}
\end{center}
\end{figure}

It is evident from figure \ref{deddynk} that $DE_{10}$ has a
regular $D_9$ subalgebra, and therefore also admits a decomposition
into $SO(9,9)$ tensors. To facilitate the comparison with $E_{10}$,
we have chosen the same labeling conventions for both Dynkin
diagrams. In order to distinguish the two decompositions, we will
in this subsection denote by $\ell^D$ the level in the decomposition
of $DE_{10}$ with respect to $D_9$ and by $\ell^E$ the level of
the $E_{10}$ decomposition. The level $\ell^E=\ell^D=0$ sectors
contain the adjoint of $\mathfrak{so}(9,9)$ and a scalar in both
decompositions. Since we tentatively identified these fields in
the $E_{10}$ decomposition with the type I fields, the two
$\mathfrak{so}(9,9)$ are naturally identified. In order to retrieve
$DE_{10}$ in $E_{10}$ all that remains to be done is to identify
the simple root $\a_0^D$ of $DE_{10}$ as a real root in the
$E_{10}$ root lattice; the subalgebra property is then a simple
consequence of Thm. 3.1 of \cite{FeNi03}.

The type I fields are (contained in) the NSNS fields which belong
to the even levels $\ell^E\in 2\ints$ of the $E_{10}$ decomposition.
The only representation at level $\ell^E = 2$ has Dynkin labels
$[001000000]$ and is generated by a highest weight vector in the
root space of
\be\label{deembed}
\a_0^D \equiv
 \a_4+2\a_5+3\a_6+4\a_7+3\a_8+2\a_9+2\a_0^E
 =: 2\a_0 + \sum_{i=1}^9 n_i \a_i
\ee
in the $E_{10}$ root lattice. This formula is easily read off
from the list of highest weight vectors (see appendix \ref{eddec}),
and it is also easy to check that $\a_0^D$ is a {\em real} root. This
root also appeared in \cite{SchnWe04} in the context of very-extended
symmetries of type I theories as a subsector of type II.
We can thus adjoin this root to the nine (common) simple roots of
$\mathfrak{so}(9,9)$ called $\a_i$ ($i=1,\ldots,9$). Using the
inner product in $E_{10}$ one verifies that their inner product
matrix is just the Cartan matrix of $DE_{10}$; hence, the ten elements 
$\a_0^D$, $\a_i$ constitute a
set of simple roots within $E_{10}$ as none of their differences
is a root. Therefore \cite{FeNi03}, they generate a subalgebra of
$E_{10}$ which is just $DE_{10}$. This could have been guessed
already by inspection of the $DE_{10}$ diagram, where the representation
at $\ell^D=1$ corresponding to the simple root $\a_0^D$ is easily
is seen to be $[001000000]$, which is also the only representation
appearing at level $\ell^E =2$ for $E_{10}$.

We note a few immediate consequences of the embedding
$DE_{10}\subset E_{10}$. First of all, if a root $\a$ is a
root of both $DE_{10}$ and $E_{10}$, and therefore belongs to
the root lattice of $DE_{10}$ embedded in the $E_{10}$ root lattice,
the multiplicity of $\a$ as root of $DE_{10}$ cannot exceed its
multiplicity as an $E_{10}$ root\footnote{In fact, inspection
of the available tables of root multiplicities suggests the
stronger inequalities \cite{Kl04}
$$
{\rm mult}_{DE_{10}} (\alpha) \leq p_8 \left(1 - \ft12 \alpha^2\right)
\leq {\rm mult}_{E_{10}}(\alpha).
$$}
\be\label{rtmulsineq}
{\rm mult}_{DE_{10}} (\alpha) \leq {\rm mult}_{E_{10}}(\alpha).
\ee

We recall that the condition for a positive $DE_{10}$ root $\a=\sum_{a=0}^9
m_a^D\a_a^D$ to contain a highest weight vector for a
$\mathfrak{so}(9,9)$ representation with Dynkin labels $[p_1\ldots
p_9]$ at level $\ell^D=m_0^D$ is \cite{DaHeNi02,NiFi03}
\be
m_i^D = - \sum_{j=1}^9 S^{-1}_{ij} p_j + S^{-1}_{i3} \ell^D,
\ee
for $i=1,\ldots,9$, where $S$ is the Cartan matrix of
$D_9$. Similarly, the condition for $E_{10}$ and the same
representation is
\be
m_i^E = - \sum_{j=1}^9 S^{-1}_{ij} p_j + S^{-1}_{i9} \ell^E.
\ee
Subtracting the two conditions, and using $\ell^E=2\ell^D$ from the
explicit form in (\ref{deembed}), we find
\be
m_i^E = m_i^D + \ell^D n_i,
\ee
in agreement with the embedding of the lattices. Consequently, any
admissible representation in the $DE_{10}$ decomposition yields
an allowed representation in the $E_{10}$ representation. The
converse can be shown to be true using the hyperbolicity of
$DE_{10}$.

Using the embedding it is also evident that for each admissible
representation of both $DE_{10}$ and $E_{10}$, the outer multiplicities
obey an inequality analogous to (\ref{rtmulsineq}), namely
\be\label{outmulsineq}
\m_{DE_{10}}(\a) \le \m_{E_{10}}(\a).
\ee
This inequality follows from the fact that all representations in
$DE_{10}$ are obtained from commutation of the $\ell^D=1$
representation and this representations is present on $\ell^E=2$
in $E_{10}$ obeying the same commutation relations. Additional
fields in $E_{10}$ can arise by taking commutators of odd level
elements into account, thereby increasing the outer multiplicity
compared to its value in $DE_{10}$ decomposition. The effect of 
these additional fields in view of (\ref{outmulsineq}) can be 
studied in the tables of appendices \ref{eddec} and \ref{deddec}.

As a ten dimensional {\sl string theory}, type I is obtained from type
IIA by gauging a combination between world-sheet parity and
space-time parity. (This breaks the Poincar\'e symmetry in ten
dimensions; also clear since the $32$ D8-branes required for the
only consistent $SO(32)$ gauge group, act as domain walls.
Actually, one should start from IIB but after compactification
they are equivalent anyway.) This parity is seen to leave all the
NSNS-fields intact and we can study (at least empirically) the
question if the type I NSNS fields are {\sl all} fields on the
even levels of $E_{10}$ decomposed with respect to
$\mathfrak{so}(9,9)$. (Algebraically, this question is equivalent
to studying the obvious subalgebra given by all even levels $\ell^E\in
2\ints$ and
asking whether it is identical to $DE_{10}$.) From the tables in
appendices \ref{eddec} and \ref{deddec}, it is apparent that there
are fields in $E_{10}$ which qualify as NSNS fields but which do
not derive from the type I subalgebra. An example of such a field
is the representation $[000000100]$ on level $\ell^E=6$
($\ell^D=3$). Assuming the validity of $E_{10}$ as a symmetry of
M-theory we can therefore learn something about the additional
degrees of freedom from studying these tables.

Finally, we note that the decomposition of $E_{10}$ under its
regular $A_9$ subalgebra, as studied in \cite{DaHeNi02,NiFi03},
can be used to establish that (the non-hyperbolic, over-extended)
$AE_{10}$ is a subalgebra of $E_{10}$ as well. This was already
shown in \cite{KlSchnWe04} for all very-extended algebras where
this subalgebra corresponded to the gravity subsector of the
generalised M-theories. The new simple root is now
\be
\a_0^A =  \a_3 + 2 \a_4 + 3 \a_5 + 4 \a_6 + 5 \a_7
    + 3 \a_8 + \a_9 + 3 \a_0^E
\ee
As in \cite{KlSchnWe04}, $AE_{10}$ is associated with the gravity 
sector here, too (see also \cite{DaHeJuNi01}). The levels containing 
fields belonging to that subalgebra are multiples of three: $\ell =3n$. 
The embedding $AE_{10}\subset E_{10}$ implies inequalities analogous to 
(\ref{rtmulsineq}) and (\ref{outmulsineq}).

\end{subsection}

\end{section}

\begin{section}{\bf \bmath $E_{10}/K(E_{10})$ \ubmath coset space dynamics}
\label{sigmamod}

\begin{subsection}{\bf \bmath The $E_{10}/K(E_{10})$ $\s$-model
at levels $\ell = 0,\pm1$\ubmath}
\setcounter{equation}{0}

In \cite{DaHeNi02} it was shown that a truncated version of the
$D=11$ supergravity equations of motion, where one retains only
the fields and their first order spatial gradients, can be mapped
onto a constrained null geodesic motion in the infinite dimensional
coset space $E_{10}(\reals)/K(E_{10})$. The detailed comparison
there was based on a level expansion in terms of the $A_9$ subalgebra
of $\lae_{10}$ up to level $\ell=3$, or alternatively
up to height 30 in the roots of $\lae_{10}$. Here, we will repeat
this analysis, but now in terms of the level expansion of $\lae_{10}$
w.r.t. its $D_9$ subalgebra, using the results of the foregoing
sections. While the $A_9$ decomposition is appropriate to the
direct reduction from eleven to one dimensions, with the $GL(10)$
global symmetry acting in the obvious way on the spatial zehnbein,
the $D_9$ decomposition is linked to the reduction of the IIA theory
from ten to one dimensions, and hence by duality also to the
type IIB theory, as we have explained. Apart from technical
differences, such as the appearance of spinor representations of $D_9$
at odd levels, the $D_9$ decomposition thus provides a different,
and more `stringy' perspective, because the group $SO(9,9,\ints)$ is the
T-duality symmetry known to arise in the compactification of the
IIA superstring to one dimension. Indeed, extra work was required
to bring the dimensionally reduced Lagrangian into an $SO(9,9)$ resp.
$\soso$ invariant form, because only the global $SL(9,\reals)\subset SO(9,9)$
and the local $SO(9) \equiv {\rm diag} \, [\soso ]$ are manifest
in the dimensional reduction. Accordingly, the $\ell =0$ sector
already contains part of the 3-form field $A_{MNP}$, whereas
in the $A_9$ decomposition it only contains the metric (zehnbein)
degrees of freedom. Here, we will analyse the $\ell=0,1$ sector,
and perform some partial checks for the $\ell=2$ sector.

For this purpose, we slightly adapt the method of \cite{DaHeNi02}:
we will not use a background metric in order to be able to avoid
having to introduce a `spinorial metric' to deal with the odd
level spinorial representations (for the $A_9$ decomposition this 
problem does not arise because $SL(10)$ does not have spinorial 
representations). Let us also emphasize once more that the 
identification of the relevant $\s$-model quantities (cf. (\ref{e10pq})
below) with the corresponding field strengths of $D=11$ supergravity
was derived via an analysis of the supersymmetry transformations,
so the comparison with the bosonic equations of motion we are about
to perform serves as an additional consistency check.

As for finite dimensional $\s$-models, we describe the bosonic degrees 
of freedom in terms of a `matrix' $\cV(t)\in E_{10}$ depending on one 
time parameter $t$. The quantity $\pt \cV {\cV}^{-1}$ then belongs to
the Lie algebra $\lae_{10}$ and can be decomposed into a connection
$\cQ$ in the maximal compact subalgebra $\mathfrak{ke}_{10}$,
and a coset component $\cP$:
\be\label{cartform}
\pt \cV \cV^{-1} = \cQ + \cP.
\ee
In order to make the grading and dilaton dependence more explicit, we
parametrize the coset element as
\be
\cV = e^{\vp T} \tilde{\cV}
\ee
by factoring out the dilaton. With this definition, the terms
at level $\ell$ in (\ref{cartform}) will be dressed with an explicit factor
$e^{\frac{\ell}{2}\vp}$. In a triangular gauge, where the $\cV$ is the
exponential of a Borel subalgebra with only contributions from levels
$\ell\geq 0$, we thus obtain
\be
\pt \cV \cV^{-1} &=& \pt\vp \cdot T +
  e^{\vp T} \big( \pt\tilde\cV \tilde\cV^{-1} \big) e^{-\vp T} \nn\\
&=&  \pt\vp \cdot T + P_{i\bj} Y^{i\bj}
     + e^{\frac12 \vp} P^{(1)}_A E^A
     + e^{\vp} P^{(2)}_{IJK} E^{IJK} + \dots
\ee
where the dots stand for higher level contributions which we
neglect, and the superscripts indicate the level for $\ell>0$.
Splitting the terms on the r.h.s. according to whether
they belong to the compact subalgebra or the coset, we get
\be
\label{e10pq}
\cQ &=& \ft12 Q_{ij} X^{ij} + \ft12 Q_{\bi\bj} X^{\bi\bj}
  + \ft12 e^{\ft{\vp}{2}} P^{(1)}_A (E_A - \CC_{A\dB}F_\dB)
  + \ft12 e^\vp P^{(2)}_{IJK} (E^{IJK} - F_{IJK}) + \dots  \nn\\
\cP &=&  \pt\vp \cdot T +  P_{i\bj} Y^{i\bj}
  + \ft12 e^{\ft{\vp}{2}} P^{(1)}_A (E_A + \CC_{A\dB}F_\dB)
  + \ft12 e^\vp P^{(2)}_{IJK} (E^{IJK} + F_{IJK}) + \dots
\ee
For any variation along the coset
\be
\d \cV \cV^{-1} =
\L \equiv \L_{i\bj} Y^{i\bj} + \ft12 \L_A (E_A + \CC_{A\dB} F_\dB)
+ \ft12 \L_{IJK} (E^{IJK} + F_{IJK}) \dots,
\ee
with $-\o(\L)=\L$, we therefore have
\be
\d \left( \pt \cV \cV^{-1} \right) =
\pt \L - [\cQ , \L] - [\cP , \L ]
\ee
Thus, the local supersymmetry variations at levels $\ell =0$ and
$\ell =1$ are identified as follows, cf.~(\ref{Libj}) and (\ref{dPaa}),
\begin{subequations}
\be
\L_{i\bj} &=& i \ve_\a \g_{i\a\b} \chi_{\bj \b} +
              i \ve_\ba \g_{\bj\ba\bb} \chi_{i \bb} \\
\L_A &=& \L_{\a\ba} =  2 i \Big[ \ve_\a\chi_\ba  -\ve_\ba\chi_\a +
 \ve_\b\g_{i\b\a}\chi_{i\ba} -\ve_\bb\g_{\bj\bb\ba}\chi_{\bj\a}\Big],
\ee
\end{subequations}
where we have rewritten the $SO(9,9)$ spinor as an $({\bf 16},{\bf 16})$ 
bispinor under $\soso$ as before.
The above formulae are convenient both for deriving the equations
of motion as well as keeping track of the dilaton dependence in
the variations of the $P$'s. For instance, we have 
\be
\d P_{i\bj} &=& D_t \L_{i\bj} \nn\\
\d \big(e^{\frac12 \vp} P_A^{(1)} E^A \big) &=& D_t \L_A E^A +
\frac12 e^{\frac12 \vp} \big[ P_A^{(1)} ( E^A - \CC^{A\dB} F_\dB)\,
,\, \L_{i\bj} Y^{i\bj} \big] + \dots
\ee

The (bosonic) null geodesic motion in $E_{10}/K(E_{10})$ is
governed by the Lagrange function \cite{DaHeNi02}
\be
\label{cosetlag}
\cL = \frac{1}{4}n^{-1} (\cP|\cP)
\ee
which gives rise to the $K(E_{10})$-covariant equations of motion
in the standard fashion:
\be
\label{coseteom}
\cD_t (n^{-1} \cP) \equiv \pt (n^{-1}\cP) - n^{-1}[\cQ,\cP] = 0.
\ee
Since we understand the Kac--Moody algebra $\lae_{10}$ at the 
very lowest levels only, we truncate the expansion (\ref{e10pq}) 
and the evaluation of (\ref{coseteom}) after some level to obtain an
approximation to the dynamics. Notice that the higher level
terms come with additional powers of the `string coupling'
$e^{\frac{1}{2}\vp}$ (recall, however, that our dilaton is different
from the standard IIA string dilaton), which would be suppressed 
for $\langle\vp\rangle <0$. In order to write the equations of
motion we explicitly covariantize with respect to $\soso$ by
replacing $\pt$ by the $\soso$ covariant $D_t$ and leaving out
the contribution from $Q_{ij}$ and $Q_{\bi\bj}$ on the r.h.s. of
(\ref{coseteom}). This results in expressions like (\ref{sosocov})
for spinor and vector indices.

The expansion of the coset Lagrange function up to level $\ell=1$
yields
\be
\cL = \frac14 n^{-1}\Big(P_{i\bj}P_{i\bj}-(\pt\vp)^2\Big) +\frac18
n^{-1} e^\vp P_{\a\ba}P_{\a\ba}\; ,
\ee
and this agrees indeed with the first line of the reduced type II
Lagrangian (\ref{typeIIlag}). Note that $P_A^{(1)}=P_{\a\ba}$ is 
to be expanded in terms of {\em odd} degree $\g$-matrices since 
all $\vp$-dependence resides in the prefactor $e^\vp$.

From (\ref{coseteom}) we can derive the $\sigma$-model equations
of motion for levels $\ell\leq 1$, and compare them with those 
of massive IIA supergravity. The bosonic equations of motion 
at levels $\ell =0,1$ that follow from (\ref{coseteom}), with
contributions up to $|\ell|=2$, read
\begin{subequations}\label{coseteoml1}
\be
\p_t \big(n^{-1}\pt \vp\big) &=& 
-\frac{1}{4}n^{-1}e^\vp P_{\a\ba} P_{\a\ba}  + \\
&&\qquad +
    \frac{1}{24}n^{-1}e^{2\vp}
    \Big(P_{ijk} P_{ijk} - 3 P_{ij\bk} P_{ij\bk}
    + 3 P_{i\bj\bk} P_{i\bj\bk} -  P_{\bi\bj\bk}
    P_{\bi\bj\bk} \Big) + \ldots \nn\\
\label{pijeom}
D_t \big(n^{-1} P_{i\bj}\big) &=& \frac{1}{4} n^{-1} e^\vp
   \g^i_{\a\b} \g^\bj_{\ba\bb} P_{\a\ba} P_{\b\bb}  - \\
 && \qquad-\frac{1}{4}n^{-1}e^{2\vp} \Big(P_{ikl}P_{\bj kl}
  -2P_{ik\bl}P_{\bj k
 \bl}+P_{i\bk\bl}P_{\bj\bk\bl}\Big)+\ldots\nn\\
 \label{lev1eom}
D_t \big(n^{-1}e^\vp P_{\a\ba}\big) &=&
  \frac12 \g^i_{\a\b} \g^\bj_{\ba\bb} n^{-1}e^\vp P_{i\bj}
   P_{\b\bb} + \\
&&\!\!\!\!\!\!\!\!\!\!\!\!\!\!\!\!\!\!\!\!\!\!\!\!\!\!\!\!\!\!\!\!\!\!\!
\!\!\!\!\!\!\!\!\!\!\!
  +\frac{1}8 n^{-1}e^{2\vp} P_{\b\bb}
    \Big( -P_{ijk} \g^{ijk}_{\a\b} \d_{\ba\bb}
    + 3P_{ij\bk} \g^{ij}_{\a\b} \g^\bk_{\ba\bb}
    - 3P_{i\bj\bk} \g^{i}_{\a\b} \g^{\bj\bk}_{\ba\bb}
 +  P_{\bi\bj\bk} \g^{\bi\bj\bk}_{\ba\bb} \d_{\a\b}\Big) +
 \dots\nn
\ee
\end{subequations}
where we have now dropped the superscripts indicating the level for 
simplicity of notation, and where, for instance, the 
last line is obtained by expanding out
$\Gamma^{IJK}_{AB} P^{(1)}_B P^{(2)}_{IJK}$ in terms of $\soso$
indices. We have included the $\ell=2$ contributions for completeness
but will only make partial use of these terms 
when relating (\ref{coseteoml1}) to the reduced
massive ${\rm IIA}$ equations of motion.

\end{subsection}

\begin{subsection}{\bf Equivalence to type II supergravity}

For the comparison with massive IIA supergravity, we first note that
the correctness of the $\ell =0$ truncation already follows from our 
rewriting of the reduced Lagrangian into $\s$-model form. In particular,
the $\soso$ covariant derivative $D_t$ takes care not only of the terms
involving the spin connection, but also the couplings of
the NSNS field $A_{mn\tten}$.

We will now demonstrate in examples that the $\s$-model equations 
(\ref{coseteoml1}) coincide with the various components of the
bosonic $D=11$ supergravity equations of motion (\ref{sugraeom})
and the Bianchi identities (\ref{Bianchi}). For the latter we will
use flat indices, which results in extra contributions from the 
spatial components of the spin connection, as in (\ref{Bianchi1}). 
For the contributions 
generated by the Romans mass $M$, we merely check the couplings against
the results of \cite{Ro86}, but not the precise numerical coefficients.

At level $\ell=0$, we deduce the following equations of motion for 
$P_{i\bj}$ and $\vp$ from our definitions (\ref{QP}), 
\begin{subequations}
\label{l0eom}
\be
\label{pijricci}
\frac12 n D_t\big(n^{-1}(P_{i\bj}+ P_{j\bi})\big) 
&=&-N^{2}\big( R_{ij}+\frac12\d_{ij}R_{10,10}
  -2e_i{}^m e_j{}^n g^{pq}F_{tmp\tten}F_{tnq\tten}\big), \\
\frac12 n D_t\big(n^{-1}(P_{i\bj}-P_{j\bi})\big) &=& -
n\pt\big(n^{-1}g^{mp} g^{nq} F_{tpq\tten}\big),\\
\label{vpricci}
n \pt\big( n^{-1}\pt\vp\big) &=& -N^{2}\big( -\frac32
  R_{10,10} -\d^{ab}R_{ab}\big).
\ee
\end{subequations}
We have separated the symmetric and anti-symmetric part of the equation
for $P_{i\bj}$ since it is clear from (\ref{QP}) that the first will
correspond to Einstein's equation of motion while the latter should
reduce to the equation for the NSNS 2-form. It is easy to check from
(\ref{sugraeom}) that they are equivalent to the vanishing of the 
equations (\ref{l0eom}) if one considers only contributions from $\ell=0$.
Note that the combination of Ricci tensors in (\ref{pijricci}) is
correct in that it cancels contributions of the form $\d_{ij}
g^{mn}g^{pq}F_{tmp\tten}F_{tnq\tten}$  from (\ref{einsteineom}) 
as required, since (\ref{pijeom}) does not have any such trace terms.
The contribution to (\ref{pijricci}) from the RR fields at $\ell=1$ 
on the r.h.s. of (\ref{pijeom}) yields, taking $G_{tmnp}$ as an example,
\be
-n^{-1}e^\vp g^{\frac12}g^{mn}g^{pq} G_{timp}G_{tjnq}
+\frac1{6}n^{-1}e^\vp g^{\frac12}
\d_{ij}g^{mn}g^{pq}g^{rs} G_{tmpr}G_{tnqs}.
\ee
(note the appearance of the redefined field strength (\ref{Gtmnp})).
This coincides indeed with the corresponding term on the r.h.s. of
Einstein's equations (\ref{einsteineom}). The other contributions work
analogously. 

To analyse the $\ell =1$ equations for $P_{\a\ba}$, we make use 
of the expansion (\ref{paa1}) (rather than (\ref{paa2})) and first 
write out the covariantizations containing the $\ell =0$ fields
\be\label{D0paa}
&& \!\!\!\!\!nD_t( n^{-1} e^\vp P_{\a\ba}) 
   - \frac12 e^\vp P_{i\bj} \big( \g^i P \g^\bj \big)_{\a\ba} = 
     n \pt ( n^{-1} e^\vp P_{\a\ba})  +  
    \frac14 e^\vp {e_i}^m \pt e_{mj}\, \big[ \g^{ij} , P \big]_{\a\ba} + \nn\\
&&  \,\,+ \frac14 e^\vp F_{tij\tten}\, \big\{ \g^{ij} , P \big\}_{\a\ba} 
    -\frac12 e^\vp {e_i}^m \pt e_{jm} \, \big( \g^{(i} P \g^{j)} \big)_{\a\ba}
   + \frac12 e^\vp F_{tij\tten} \, \big( \g^{[i} P \g^{j]} \big)_{\a\ba}
\ee
The structure of this equation explains the necessity of the factor 
$g^{\frac14}$ which we first encountered in (\ref{paa1}): contracting
${e_{(i}}^m \pt e_{mj)}$ with the trace term coming from
\be
\g^i \g^{(p)} \g^j = (-1)^p \d^{ij} \g^{(p)} + \dots
\ee
produces a contribution which either cancels the derivative
$\pt g^{1/4}$ in $\pt P_{\a\ba}$ for even $p$ (and thus gives the 
Bianchi identities) or enhances it to the desired $\pt g^{1/2}$ 
required by (\ref{typeIIred}) for odd $p$ (and thus the equations 
of motion). Furthermore, the other terms involving the time derivative
of the neunbein $e_{im}$ conspire to give either derivatives of 
the curved `momenta' with upper indices for odd $p$, corresponding to 
equations of motion, or with lower indices, corresponding to Bianchi 
identities. Note also the occurrences of the factors $n^{-1}$ 
and $e^\vp$ in agreement with our calculations in section 
\ref{so99sec} leading to (\ref{paa1}): For even $p$, all $n$ and $\vp$
dependence in the derivative cancels in addition to the disappearance
of the $g^{\frac14}$.

It is quite remarkable how the various $\gamma$-matrix structures
in (\ref{D0paa}) conspire not only to give the correct factors
of the metric determinant, but also the correct contributions to
the equations of motion. For instance, with a little more $\g$-matrix 
algebra one can now check that the terms in (\ref{D0paa}) involving 
the NSNS field strength $F_{tij\tten}$ likewise reduce to the 
corresponding terms obtained by writing out the relevant components 
of the equations of motion (\ref{gaugeeom}). Evaluating 
(\ref{D0paa}) for the term containing $\g^{(3)}$ and using the 
coefficients from (\ref{paa1}) we find up to $\ell=1$
\be\label{l1p3eom}
\pt\big( n^{-1}e^\vp g^{\frac12} g^{m_1n_1}g^{m_2n_2}g^{m_3n_3}
G_{tn_1n_2n_3}\big) = \frac1{24}
g^{\frac12}\e^{m_1m_2m_3s_1s_2r_1r_2r_3r_4} F_{tr_1r_2\tten}
F_{s_1s_2s_3s_4}.
\ee
(The appearance of $F_{s_1s_2s_3s_4}$ rather than $G_{s_1s_2s_3s_4}$
here is explained by having reconverted back to curved indices.)
This is in precise agreement with (\ref{gaugeeom}) evaluated for 
$NPQ=m_1m_2m_3$. Actually, this evaluation produces two contributions 
on the r.h.s.: one with $t$ and $\tten$ in the same field strength, and 
one where the $t$ and $\tten$ are in the two different field strength 
factors. In the latter, $F_{mnp\tten}$ corresponds to gradients of the NSNS 
two-form, and hence to a the contribution coming from the level 2 field 
$P_{IJK}$ from the r.h.s. of the $\ell=1$ equation of motion
(\ref{lev1eom}); and the contribution to (\ref{gaugeeom})
has the correct structure to agree with the $\s$-model. 

Evaluating  eq.~(\ref{D0paa}) for the  term $\g^{(p)}$ with $p=1$ gives
\be
\pt\big(n^{-1} e^\vp g^{\frac12} g^{mn}F_{tn}\big) = 2
n^{-1} g^{\frac12}e^\vp g^{mn} g^{pq} g^{rs} G_{tnpr} F_{tqs\tten}.
\ee
This can be shown to be equivalent to the equation of motion of the
Kaluza--Klein vector by reducing the eleven dimensional
Einstein equation (\ref{einsteineom}) to the type ${\rm IIA}$ equation.

For $p=4$ eq.~(\ref{D0paa}) yields 
\be\label{DtF}
\pt G_{mnpq} = 6 \pt F_{[mn} F_{pq]t\tten},
\ee
From (\ref{Bianchi1}) we see that this is indeed the correct form of 
the Bianchi identity, because the r.h.s. of (\ref{DtF}) is just the
second term on the r.h.s. of (\ref{Bianchi1}). Similarly, the first 
term on the r.h.s.of (\ref{Bianchi1}) is reproduced by the level 
$\ell=2$ contribution in (\ref{lev1eom}).
 
The equation for the field strength of the Kaluza--Klein vector
($p=2$) is
\be
\pt F_{np} = -8 M \pt F_{tnp\tten} \; ,
\ee
where $M$ is the Romans mass from (\ref{paa1}); for $M=0$ it reduces
to the Bianchi identity for the Kaluza Klein vector. 

Finally, the equation of the Romans parameter is
\be
\pt M  = 0
\ee
which makes it to a constant parameter of the theory, as
required by \cite{Ro86}.

The contributions of $M$ to the equations of 
$\vp$ and $P_{i\bj}$ from (\ref{coseteoml1}) work out correctly as well. 
For example, for $\vp$ we get a
quadratic mass contribution of the form 
\be
n\pt(n^{-1} \pt\vp) = -4 M^2 g^{\frac12} n^2 e^{-\vp} + \dots
\ee
whose structure agrees with \cite{Ro86}. Similarly, our theory also
reproduces the quadratic mass contribution to the equation of
$P_{i\bj}$ and therefore to the Einstein equation. 

In conclusion, the $\s$-model on $E_{10}/K(E_{10})$ correctly
reproduces the bosonic equations of motion of massive {\rm IIA}
supergravity in this truncation.

\end{subsection}

\end{section}

\begin{section}{\bf Discussion}
\setcounter{equation}{0}

The fact that the reduction of $N=1$, $D=11$ supergravity to one
(time-like) dimension with a mass term in the reduced theory
admits a local $\soso$ invariance was shown to be in perfect 
agreement with a $\s$-model on $E_{10}/K(E_{10})$, if one restricts 
to the bosonic sector and truncates at $|\ell|=1$ in the decomposition 
of $E_{10}$ under its $D_9$ subalgebra. Our partial checks of the 
$\ell=2$ sector containing the spatial gradients of the NSNS fields
indicate that the agreement persists to higher levels as well. 
Our analysis also incorporates the fermionic degrees of freedom, 
which could be fitted into irreducible representations of $\soso$, 
such that the resulting theory for the type I theory was locally 
supersymmetric to linear fermion order. Adding the RR fields from 
level $\ell=1$ in $E_{10}$ to extend to the full (massive) type II 
theory was accompanied by including the necessary second
set of chiral fermionic fields, and we 
noted that the resulting Lagrangian was not completely supersymmetric 
any more. From the $\s$-model viewpoint this is clearly related to the 
need of introducing all the infinitely many other fields contained in 
$E_{10}$ which can be generated by commutators of $\ell=1$ fields. 
In order to obtain a fully supersymmetric model one would need to 
also include infinitely many new fermionic degrees of freedom,
extending the finite dimensional spinor representations of $\soso$
to a single irreducible infinite dimensional spinor representation 
of $K(E_{10}) $\footnote{The `wave function of the universe', alias
modular form of \cite{BrGaHe04,PioWal02}, would thus have to satisfy 
a {\em linear} supersymmetry constraint rather than a generalized 
Laplace equation on $E_{10}(\ints)\!\setminus\! E_{10}/K(E_{10})$}. 
Constructing such a representation is one of the tantalizing 
challenges in the current framework. 

As we mentioned in the introduction, the $D_9$ decomposition is
thought to be `stringier' than the one in terms of $SL(10)$
representations because $SO(9,9,\ints)$ is directly identified as
a string T-duality group. One would thus hope that the decomposition
of $E_{10}$ under its $D_9$ subalgebra would eventually provide
more insight into the higher level fields, partly because of the
analogy with NSNS and RR fields on even and odd levels. However,
we have not been able so far to detect traces of massive string
states in the tables of appendix \ref{eddec}. The special role 
assigned to the $10$-th spatial direction necessary for the 
$SO(9,9)$ decomposition makes the gradient conjecture of 
\cite{DaHeNi02} more elusive. The relevant representations are 
now $18$ component derivatives, which we tentatively associated 
with left and right derivatives along the remaining nine spatial 
directions, modulo a tracelessness condition. The `gradient' w.r.t. 
the dilaton direction remains mysterious in this set-up.

Finally, we recall that $N=1, D=11$ supergravity can be viewed as 
a strong coupling limit of type IIA supergravity via the
identification \cite{Wi95} 
\be
R_{11} = (\a')^{1/2} g_s^{2/3}, \qquad \kappa_{11} = (\a')^{9/2}
g_s^3,
\ee
where $R_{11}$ is the radius of the compactified 10-th spatial 
dimension. Thus the small tension limit $\a'\rightarrow\infty$ at 
fixed string coupling $g_s = e^{2\langle \phi \rangle}$ corresponds 
to the decompactification limit $R_{11}\rightarrow\infty$. The 
level decomposition with the power of the dilaton as the grading
resembles an expansion in powers of the string coupling constant, 
even though the algebraic dilaton $\vp$ is different from the string 
dilaton $\phi$. This is another point which deserves further study.
\\{}\\{}\\

\noindent
{\bf Acknowledgments:}
We would like to thank Thibault Damour, Peter Goddard and Marc Henneaux
for discussions and comments.
AK would like to thank the Albert-Einstein-Institut, Golm, for its
hospitality and support during a number of visits while this work was
conceived, and is also grateful to the Institute for Advanced Study,
Princeton, for hospitality and support while part of this work was
carried out. The research of AK was supported by the Studienstiftung des
deutschen Volkes.

\end{section}

\newpage

\appendix

\setcounter{equation}{0}

\begin{section}{\bf Gamma matrix conventions}
\label{gconv}
We here summarize our conventions for the three types of $\g$-matrices
needed in this paper, namely for $SO(9)$, $SO(1,10)$, and $SO(9,9)$,
respectively.

\subsection{$SO(9)$}
The $SO(9)$ $\g$-matrices $\g^i$ are real symmetric $16\times 16$
matrices obeying
\be
\{ \g^i , \g^j \} = 2 \d^{ij} \;\;\; , \quad
\g^{1} \cdots \g^{9} = \boldone
\ee
where $\boldone$ is the $16\times 16$ unit matrix. A complete set
of $16\times 16$ matrices is is obtained by forming the standard
antisymmetric combinations $\g^{ij} \equiv \g^{[i} \g^{j]}$,
etc. Among these, $\boldone$, $\g^i$, and $\g^{ijkl}$ are symmetric,
while $\g^{ij}$ and $\g^{ijk}$ are antisymmetric. We thus have the
completeness relation
\be
T_{\a\b}&=&\frac{1}{16}\d_{\a\b} T_{\g\g}
+\frac{1}{16}\g^i_{\a\b}
\g^i_{\g\d}T_{\d\g}
-\frac{1}{2!\cdot 16}\g^{ij}_{\a\b}
\g^{ij}_{\g\d}T_{\d\g}\nonumber\\
&&-\frac{1}{3!\cdot 16}\g^{ijk}_{\a\b}
\g^{ijk}_{\g\d}T_{\d\g}
+\frac{1}{4!\cdot 16}\g^{ijkl}_{\a\b}
\g^{ijkl}_{\g\d}T_{\d\g} \quad ,
\ee
where $T_{\a\b}$ is any real $16\times 16$ matrix. When we are dealing
with $SO(9)\times SO(9)$, the second set of $SO(9)$ $\g$-matrices
is labeled with barred vector and spinor indices, i.e. we write
$\g^\bi_{\ba\bb}$.

Note also that the matrices $\g^{ij}$ and $\g^i$ together generate
the non-compact group $SO(1,9)$.

\subsection{$SO(1,10)$}
The $SO(1,10)$ $\g$-matrices will be designated by $\G^A$ (with $A,B,... =
0,1,\dots, 10$). They obey
\be
\left\{ \G^A,\G^B\right\}=2\eta^{AB}
\ee
with the ``mostly positive'' metric $\eta^{AB}$. In a Majorana
representation where $\G^0 = \CC$ (the charge conjugation matrix)
we can express them directly in terms of the $SO(9)$ $\g$-matrices
introduced above
\be
\G^i&=&\left(\begin{array}{cc}0&\g^i\\\g^i&0\end{array}\right)
= \s_1 \otimes \g^i\\
\G^{10}&=&\left(\begin{array}{cc}\boldone&0
\\0&-\boldone\end{array}\right) = \s_3 \otimes \boldone\\
\G^0&=&\left(\begin{array}{cc}0&-\boldone\\ \boldone&0\end{array}\right)
= \e \otimes \boldone = \CC
\ee
with the standard $\s$-matrices
\be
\s_1=\left(\begin{array}{cc}0&1\\1&0\end{array}\right)\qquad
\e=\left(\begin{array}{cc}0&-1\\1&0\end{array}\right) \qquad
\s_3=\left(\begin{array}{cc}1&0\\0&-1\end{array}\right)
\ee

\subsection{$SO(9,9)$}
The $SO(9,9)$ $\g$-matrices are real $512\times 512$ matrices, and are
conveniently written as direct products of the $SO(9)$ $\g$-matrices
given above. To distinguish the $SO(9,9)$ $\g$-matrices from the
previous matrices, we put a hat on them. We have
\be
\wG^{i} = \s_1\otimes\g^i\otimes\boldone\;\; , \quad
\wG^{\bi} = \e\otimes\boldone\otimes\g^i
\ee
with $(i,\bi)=I\in \{1,\dots ,9, \bar 1,\dots , \bar 9\} \equiv
\{1,\dots, 18\}$. One easily checks that with (\ref{eta})
\be\label{cliff99}
\left\{ \wG^I,\wG^J\right\}=2\eta^{IJ}
\ee
as required. The representations relevant for the decomposition
of $E_{10}$ are the {\it chiral} eigenspinors of
\be\label{G*}
\wG^{*}:=\wG^1 \cdots \wG^9 \wG^{\bar 1} \cdots \wG^{\bar 9}
  =  \s_3\otimes\boldone \otimes\boldone.
\ee
The two representations will be denoted by $\bf{256}_s$ and $\bf{256}_c$,
respectively. We also need the charge conjugation matrix of $SO(9,9)$,
which in this notation is given by
\be
\wCC=\e\otimes\boldone\otimes\boldone.
\ee
and obeys
\be
\wCC\, \wG^I \wCC^{-1} = - (\wG^I)^T \quad , \qquad
\wCC\, \wG^* + \wG^* \wCC = 0.
\ee
Our representation of the $SO(9,9)$ $\g$-matrices is adapted to chiral
spinor representations. In terms of $256\times 256$ `$\s$-matrices'
{\it \`a la} Weyl-van der Waerden, and using undotted indices
$A,B,...\in\{1,...,256\}$ and dotted indices $\dA,\dB,...\in\{1,...,256\}$
they can can be written as
\be
\wG^I=\left(\begin{array}{cc}0&\S^I_{A\dB}\\
\bar{\S}^{I}_{\dA B} &0\end{array}\right);\qquad\qquad
\wCC=\left(\begin{array}{cc}0&\CC_{A\dB}\\
\CCb_{\dA B}&0\end{array}\right)
\ee
where $\CC_{A\dB} = \d_{A\dB}$ with our choice of basis, 
$\bar{\S}^i = \big( \S^i \big)^T$ and
$\bar{\S}^\bi = - \big( \S^\bi \big)^T$. In $SO(9)\times SO(9)$ 
bispinor notation we have
\be
\S^i_{\a\ba,\b\bb} = \g^i_{\a\b}\d_{\ba\bb} = + \bar{\S}^i_{\a\ba,\b\bb}
\quad , \qquad
\S^\bi_{\a\ba,\b\bb} = \d_{\a\b}\g^\bi_{\ba\bb} = - \bar{\S}^\bi_{\a\ba,\b\bb}.
\ee
The Clifford relation (\ref{cliff99}) then reads
\be
\bar{\S}^I_{\dA C} \S^{J}_{C \dB} + \bar{\S}^J_{\dA C} \S^{I}_{C \dB}
&=& 2 \eta^{IJ} \d_{\dA\dB},\\
\S^I_{A\dC} \bar{\S}^{J}_{\dC B} + \S^J_{A\dC} \bar{\S}^{I}_{\dC B}
&=& 2 \eta^{IJ} \d_{AB}.
\ee
Furthermore, we define
\be
\S^{IJ}_{AB} := \S^{[I}_{A\dC} \bar{\S}^{J]}_{\dC B}
\quad , \qquad
\bar{\S}^{IJ}_{\dA\dB} :=
\bar{\S}^{[I}_{\dA C} {\S}^{J]}_{C \dB}.
\ee
The product $\bf{256}_s \otimes \bf{256}_c$ contains the singlet
$\CC_{A\dB} := \d_{A\dB}$ which can be used to convert dotted into 
undotted indices, and vice versa. We have for example
\be
\S^{IJ} + \CC \big(\bar\S^{IJ}\big)^T \CC^{-1} = 0.
\ee

\end{section}

\begin{section}{\bf \bmath Decomposition of $E_{10}$ with respect to $D_9$
\ubmath}
\label{eddec}

In this appendix we present the first few complete levels of the
decomposition of $E_{10}$ into representations of its $D_9$
subalgebra indicated in figure \ref{eddynk}. The $GL(1)$ charge in
our conventions is always equal to $\ell/2$ where $\ell$ is the
level in the table below, {\sl i.e.} the number of times the root
$\a_{0}$ appears in the decomposition of a root
\be
\a = \ell \a_0 + \sum_{i=1}^9 m_i\a_i \; .\nn
\ee
Furthermore, $\mu$ denotes the outer multiplicity with which a
given $D_9$ representation occurs. For the decomposition technique see
\cite{DaHeNi02,NiFi03,KlSchnWe04,We03a}. This computation can be
carried out much further.

{\small

\begin{longtable}{|c|c|c|c|c|c|r|}
\hline
$\ell$&\mbox{Dynkin $[p_s]$}&$\alpha\in E_{10}$&$\alpha^2$&\mbox{mult}$(\alpha)$&$\mu$&\mbox{dim}\\
\hline\hline
\endhead
1&[000000010]&(0,0,0,0,0,0,0,0,0,1)&2&1&1&256\\
\hline
2&[001000000]&(0,0,0,1,2,3,4,3,2,2)&2&1&1&816\\
2&[100000000]&(0,1,2,3,4,5,6,4,3,2)&0&8&0&18\\
\hline
3&[100000010]&(0,1,2,3,4,5,6,4,3,3)&2&1&1&4352\\
3&[000000001]&(1,2,3,4,5,6,7,5,3,3)&0&8&0&256\\
\hline
4&[000001000]&(1,2,3,4,5,6,8,6,4,4)&2&1&1&18564\\
4&[101000000]&(0,1,2,4,6,8,10,7,5,4)&2&1&1&11475\\
4&[000100000]&(1,2,3,4,6,8,10,7,5,4)&0&8&0&3060\\
4&[200000000]&(0,2,4,6,8,10,12,8,6,4)&0&8&0&170\\
4&[010000000]&(1,2,4,6,8,10,12,8,6,4)&-2&44&1&153\\
4&[000000000]&(2,4,6,8,10,12,14,9,7,4)&-4&192&0&1\\
\hline
5&[001000001]&(1,2,3,5,7,9,11,8,5,5)&2&1&1&169728\\
5&[200000010]&(0,2,4,6,8,10,12,8,6,5)&2&1&1&39168\\
5&[010000010]&(1,2,4,6,8,10,12,8,6,5)&0&8&1&34560\\
5&[100000001]&(1,3,5,7,9,11,13,9,6,5)&-2&44&1&4352\\
5&[000000010]&(2,4,6,8,10,12,14,9,7,5)&-4&192&1&256\\
\hline
6&[010010000]&(1,2,4,6,8,11,14,10,7,6)&2&1&1&930240\\
6&[100000011]&(1,3,5,7,9,11,13,9,6,6)&2&1&1&707200\\
6&[100001000]&(1,3,5,7,9,11,14,10,7,6)&0&8&1&293760\\
6&[201000000]&(0,2,4,7,10,13,16,11,8,6)&2&1&1&90288\\
6&[000000020]&(2,4,6,8,10,12,14,9,7,6)&0&8&0&24310\\
6&[000000002]&(2,4,6,8,10,12,14,10,6,6)&0&8&0&24310\\
6&[011000000]&(1,2,4,7,10,13,16,11,8,6)&0&8&1&67830\\
6&[000000100]&(2,4,6,8,10,12,14,10,7,6)&-2&44&2&31824\\
6&[100100000]&(1,3,5,7,10,13,16,11,8,6)&-2&44&2&45696\\
6&[000010000]&(2,4,6,8,10,13,16,11,8,6)&-4&192&1&8568\\
6&[300000000]&(0,3,6,9,12,15,18,12,9,6)&0&8&0&1122\\
6&[110000000]&(1,3,6,9,12,15,18,12,9,6)&-4&192&2&1920\\
6&[001000000]&(2,4,6,9,12,15,18,12,9,6)&-6&727&3&816\\
6&[100000000]&(2,5,8,11,14,17,20,13,10,6)&-8&2472&1&18\\
\hline
7&[100100010]&(1,3,5,7,10,13,16,11,8,7)&2&1&1&8186112\\
7&[000001001]&(2,4,6,8,10,12,15,11,7,7)&2&1&1&2558976\\
7&[020000001]&(1,2,5,8,11,14,17,12,8,7)&2&1&1&1740800\\
7&[000010010]&(2,4,6,8,10,13,16,11,8,7)&0&8&1&1410048\\
7&[101000001]&(1,3,5,8,11,14,17,12,8,7)&0&8&2&2276352\\
7&[000100001]&(2,4,6,8,11,14,17,12,8,7)&-2&44&2&574464\\
7&[300000010]&(0,3,6,9,12,15,18,12,9,7)&2&1&1&248064\\
7&[110000010]&(1,3,6,9,12,15,18,12,9,7)&-2&44&3&413440\\
7&[001000010]&(2,4,6,9,12,15,18,12,9,7)&-4&192&4&169728\\
7&[200000001]&(1,4,7,10,13,16,19,13,9,7)&-4&192&3&39168\\
7&[010000001]&(2,4,7,10,13,16,19,13,9,7)&-6&727&6&34560\\
7&[100000010]&(2,5,8,11,14,17,20,13,10,7)&-8&2472&6&4352\\
7&[000000001]&(3,6,9,12,15,18,21,14,10,7)&-10&7749&4&256\\
\hline
8&[000101000]&(2,4,6,8,11,14,18,13,9,8)&2&1&1&26686260\\
8&[110000100]&(1,3,6,9,12,15,18,13,9,8)&2&1&1&43084800\\
8&[001000020]&(2,4,6,9,12,15,18,12,9,8)&2&1&1&13579566\\
8&[001000002]&(2,4,6,9,12,15,18,13,8,8)&2&1&1&13579566\\
8&[001000100]&(2,4,6,9,12,15,18,13,9,8)&0&8&1&17005950\\
8&[101100000]&(1,3,5,8,12,16,20,14,10,8)&2&1&1&13590225\\
8&[110010000]&(1,3,6,9,12,16,20,14,10,8)&0&8&2&10340352\\
8&[000200000]&(2,4,6,8,12,16,20,14,10,8)&0&8&0&2174436\\
8&[200000011]&(1,4,7,10,13,16,19,13,9,8)&0&8&2&6096948\\
8&[001010000]&(2,4,6,9,12,16,20,14,10,8)&-2&44&3&3837240\\
8&[010000011]&(2,4,7,10,13,16,19,13,9,8)&-2&44&4&5290740\\
8&[200001000]&(1,4,7,10,13,16,20,14,10,8)&-2&44&3&2494206\\
8&[010001000]&(2,4,7,10,13,16,20,14,10,8)&-4&192&5&2131800\\
8&[030000000]&(1,2,6,10,14,18,22,15,11,8)&2&1&1&261800\\
8&[301000000]&(0,3,6,10,14,18,22,15,11,8)&2&1&1&516800\\
8&[100000020]&(2,5,8,11,14,17,20,13,10,8)&-4&192&3&393822\\
8&[100000002]&(2,5,8,11,14,17,20,14,9,8)&-4&192&3&393822\\
8&[111000000]&(1,3,6,10,14,18,22,15,11,8)&-2&44&3&709632\\
8&[100000100]&(2,5,8,11,14,17,20,14,10,8)&-6&727&9&510510\\
8&[200100000]&(1,4,7,10,14,18,22,15,11,8)&-4&192&4&373065\\
8&[002000000]&(2,4,6,10,14,18,22,15,11,8)&-4&192&2&188955\\
8&[010100000]&(2,4,7,10,14,18,22,15,11,8)&-6&727&9&302328\\
8&[100010000]&(2,5,8,11,14,18,22,15,11,8)&-8&2472&10&132600\\
8&[000000011]&(3,6,9,12,15,18,21,14,10,8)&-8&2472&7&43758\\
8&[000001000]&(3,6,9,12,15,18,22,15,11,8)&-10&7749&10&18564\\
8&[400000000]&(0,4,8,12,16,20,24,16,12,8)&0&8&0&5814\\
8&[210000000]&(1,4,8,12,16,20,24,16,12,8)&-6&726&5&14212\\
8&[020000000]&(2,4,8,12,16,20,24,16,12,8)&-8&2472&4&8550\\
8&[101000000]&(2,5,8,12,16,20,24,16,12,8)&-10&7747&14&11475\\
8&[000100000]&(3,6,9,12,16,20,24,16,12,8)&-12&22725&9&3060\\
8&[200000000]&(2,6,10,14,18,22,26,17,13,8)&-12&22712&4&170\\
8&[010000000]&(3,6,10,14,18,22,26,17,13,8)&-14&63085&11&153\\
8&[000000000]&(4,8,12,16,20,24,28,18,14,8)&-16&167133&1&1\\
\hline
9&[010001010]&(2,4,7,10,13,16,20,14,10,9)&2&1&1&266342400\\
9&[001100001]&(2,4,6,9,13,17,21,15,10,9)&2&1&1&177365760\\
9&[200010001]&(1,4,7,10,13,17,21,15,10,9)&2&1&1&167443200\\
9&[010010001]&(2,4,7,10,13,17,21,15,10,9)&0&8&2&138498048\\
9&[111000010]&(1,3,6,10,14,18,22,15,11,9)&2&1&1&127918336\\
9&[100000012]&(2,5,8,11,14,17,20,14,9,9)&2&1&1&47297536\\
9&[100000110]&(2,5,8,11,14,17,20,14,10,9)&0&8&2&52093440\\
9&[002000010]&(2,4,6,10,14,18,22,15,11,9)&0&8&1&33196800\\
9&[200100010]&(1,4,7,10,14,18,22,15,11,9)&0&8&2&64204800\\
9&[100001001]&(2,5,8,11,14,17,21,15,10,9)&-2&44&4&38697984\\
9&[010100010]&(2,4,7,10,14,18,22,15,11,9)&-2&44&5&51270912\\
9&[100010010]&(2,5,8,11,14,18,22,15,11,9)&-4&192&8&20837376\\
9&[120000001]&(1,3,7,11,15,19,23,16,11,9)&0&8&2&16450560\\
9&[201000001]&(1,4,7,11,15,19,23,16,11,9)&-2&44&5&17199104\\
9&[000000021]&(3,6,9,12,15,18,21,14,10,9)&-2&44&2&3055104\\
9&[000000003]&(3,6,9,12,15,18,21,15,9,9)&0&8&0&1244672\\
9&[011000001]&(2,4,7,11,15,19,23,16,11,9)&-4&192&8&12729600\\
9&[000000101]&(3,6,9,12,15,18,21,15,10,9)&-4&192&4&3394560\\
9&[100100001]&(2,5,8,11,15,19,23,16,11,9)&-6&727&15&8186112\\
9&[000001010]&(3,6,9,12,15,18,22,15,11,9)&-6&727&8&2558976\\
9&[400000010]&(0,4,8,12,16,20,24,16,12,9)&2&1&1&1240320\\
9&[210000010]&(1,4,8,12,16,20,24,16,12,9)&-4&192&7&2937600\\
9&[000010001]&(3,6,9,12,15,19,23,16,11,9)&-8&2472&13&1410048\\
9&[020000010]&(2,4,8,12,16,20,24,16,12,9)&-6&727&10&1740800\\
9&[101000010]&(2,5,8,12,16,20,24,16,12,9)&-8&2472&24&2276352\\
9&[000100010]&(3,6,9,12,16,20,24,16,12,9)&-10&7749&22&574464\\
9&[300000001]&(1,5,9,13,17,21,25,17,12,9)&-6&726&6&248064\\
9&[110000001]&(2,5,9,13,17,21,25,17,12,9)&-10&7747&30&413440\\
9&[001000001]&(3,6,9,13,17,21,25,17,12,9)&-12&22725&31&169728\\
9&[200000010]&(2,6,10,14,18,22,26,17,13,9)&-12&22712&23&39168\\
9&[010000010]&(3,6,10,14,18,22,26,17,13,9)&-14&63085&39&34560\\
9&[100000001]&(3,7,11,15,19,23,27,18,13,9)&-16&167116&35&4352\\
9&[000000010]&(4,8,12,16,20,24,28,18,14,9)&-18&425227&18&256\\
\hline
10&[100010100]&(2,5,8,11,14,18,22,16,11,10)&2&1&1&1522105200\\
10&[011000011]&(2,4,7,11,15,19,23,16,11,10)&2&1&1&1655413760\\
10&[010110000]&(2,4,7,10,14,19,24,17,12,10)&2&1&1&748261800\\
10&[100100011]&(2,5,8,11,15,19,23,16,11,10)&0&8&3&1024287264\\
10&[201001000]&(1,4,7,11,15,19,24,17,12,10)&2&1&1&848300544\\
10&[000001002]&(3,6,9,12,15,18,22,16,10,10)&2&1&1&150760896\\
10&[000001020]&(3,6,9,12,15,18,22,15,11,10)&2&1&1&150760896\\
10&[011001000]&(2,4,7,11,15,19,24,17,12,10)&0&8&2&613958400\\
10&[000001100]&(3,6,9,12,15,18,22,16,11,10)&0&8&1&154187280\\
10&[100020000]&(2,5,8,11,14,19,24,17,12,10)&0&8&1&195699240\\
10&[100101000]&(2,5,8,11,15,19,24,17,12,10)&-2&44&5&359165664\\
10&[210000002]&(1,4,8,12,16,20,24,17,11,10)&2&1&1&232792560\\
10&[210000020]&(1,4,8,12,16,20,24,16,12,10)&2&1&1&232792560\\
10&[020000002]&(2,4,8,12,16,20,24,17,11,10)&0&8&1&136334016\\
10&[020000020]&(2,4,8,12,16,20,24,16,12,10)&0&8&1&136334016\\
10&[000010011]&(3,6,9,12,15,19,23,16,11,10)&-2&44&4&164466432\\
10&[210000100]&(1,4,8,12,16,20,24,17,12,10)&0&8&2&293862816\\
10&[101000002]&(2,5,8,12,16,20,24,17,11,10)&-2&44&5&174888350\\
10&[101000020]&(2,5,8,12,16,20,24,16,12,10)&-2&44&5&174888350\\
10&[020000100]&(2,4,8,12,16,20,24,17,12,10)&-2&44&5&171348100\\
10&[101000100]&(2,5,8,12,16,20,24,17,12,10)&-4&192&11&217443600\\
10&[120100000]&(1,3,7,11,16,21,26,18,13,10)&2&1&1&106419456\\
10&[003000000]&(2,4,6,11,16,21,26,18,13,10)&2&1&1&19969950\\
10&[000011000]&(3,6,9,12,15,19,24,17,12,10)&-4&192&5&50605056\\
10&[201100000]&(1,4,7,11,16,21,26,18,13,10)&0&8&2&95178240\\
10&[000100020]&(3,6,9,12,16,20,24,16,12,10)&-4&192&5&42401502\\
10&[011100000]&(2,4,7,11,16,21,26,18,13,10)&-2&44&4&66853248\\
10&[000100002]&(3,6,9,12,16,20,24,17,11,10)&-4&192&5&42401502\\
10&[000100100]&(3,6,9,12,16,20,24,17,12,10)&-6&727&14&51395760\\
10&[210010000]&(1,4,8,12,16,21,26,18,13,10)&-2&44&5&69227298\\
10&[020010000]&(2,4,8,12,16,21,26,18,13,10)&-4&192&6&39673920\\
10&[100200000]&(2,5,8,11,16,21,26,18,13,10)&-4&192&5&28139760\\
10&[300000011]&(1,5,9,13,17,21,25,17,12,10)&-2&44&4&37209600\\
10&[101010000]&(2,5,8,12,16,21,26,18,13,10)&-6&727&19&47837592\\
10&[110000011]&(2,5,9,13,17,21,25,17,12,10)&-6&727&24&60648588\\
10&[001000011]&(3,6,9,13,17,21,25,17,12,10)&-8&2472&28&24186240\\
10&[300001000]&(1,5,9,13,17,21,26,18,13,10)&-4&192&6&15049216\\
10&[110001000]&(2,5,9,13,17,21,26,18,13,10)&-8&2472&31&24066900\\
10&[000110000]&(3,6,9,12,16,21,26,18,13,10)&-8&2472&14&9883800\\
10&[001001000]&(3,6,9,13,17,21,26,18,13,10)&-10&7749&36&9302400\\
10&[130000000]&(1,3,8,13,18,23,28,19,14,10)&0&8&1&2154240\\
10&[401000000]&(0,4,8,13,18,23,28,19,14,10)&2&1&1&2386800\\
10&[200000002]&(2,6,10,14,18,22,26,18,12,10)&-8&2472&14&3401190\\
10&[200000020]&(2,6,10,14,18,22,26,17,13,10)&-8&2472&14&3401190\\
10&[211000000]&(1,4,8,13,18,23,28,19,14,10)&-4&192&7&4558176\\
10&[021000000]&(2,4,8,13,18,23,28,19,14,10)&-6&727&10&2474010\\
10&[200000100]&(2,6,10,14,18,22,26,18,13,10)&-10&7747&37&4377296\\
10&[010000002]&(3,6,10,14,18,22,26,18,12,10)&-10&7747&27&2956096\\
10&[010000020]&(3,6,10,14,18,22,26,17,13,10)&-10&7749&29&2956096\\
10&[102000000]&(2,5,8,13,18,23,28,19,14,10)&-8&2472&14&2217072\\
10&[300100000]&(1,5,9,13,18,23,28,19,14,10)&-6&726&9&2188800\\
10&[010000100]&(3,6,10,14,18,22,26,18,13,10)&-12&22725&56&3779100\\
10&[110100000]&(2,5,9,13,18,23,28,19,14,10)&-10&7747&42&3281850\\
10&[001100000]&(3,6,9,13,18,23,28,19,14,10)&-12&22725&37&1116288\\
10&[200010000]&(2,6,10,14,18,23,28,19,14,10)&-12&22712&41&1108536\\
10&[010010000]&(3,6,10,14,18,23,28,19,14,10)&-14&63085&74&930240\\
10&[100000011]&(3,7,11,15,19,23,27,18,13,10)&-14&63085&75&707200\\
10&[100001000]&(3,7,11,15,19,23,28,19,14,10)&-16&167116&86&293760\\
10&[500000000]&(0,5,10,15,20,25,30,20,15,10)&0&8&0&25194\\
10&[310000000]&(1,5,10,15,20,25,30,20,15,10)&-8&2464&8&78336\\
10&[120000000]&(2,5,10,15,20,25,30,20,15,10)&-12&22712&24&84150\\
10&[201000000]&(2,6,10,15,20,25,30,20,15,10)&-14&63020&49&90288\\
10&[011000000]&(3,6,10,15,20,25,30,20,15,10)&-16&167099&63&67830\\
10&[000000002]&(4,8,12,16,20,24,28,19,13,10)&-16&167116&25&24310\\
10&[000000020]&(4,8,12,16,20,24,28,18,14,10)&-16&167133&26&24310\\
10&[000000100]&(4,8,12,16,20,24,28,19,14,10)&-18&425227&63&31824\\
10&[100100000]&(3,7,11,15,20,25,30,20,15,10)&-18&425156&96&45696\\
10&[000010000]&(4,8,12,16,20,25,30,20,15,10)&-20&1044218&60&8568\\
10&[300000000]&(2,7,12,17,22,27,32,21,16,10)&-16&166840&15&1122\\
10&[110000000]&(3,7,12,17,22,27,32,21,16,10)&-20&1043926&66&1920\\
10&[001000000]&(4,8,12,17,22,27,32,21,16,10)&-22&2485020&66&816\\
10&[100000000]&(4,9,14,19,24,29,34,22,17,10)&-24&5750072&27&18\\
\hline
\end{longtable}
}
\end{section}

\begin{section}{\bf \bmath Decomposition of $DE_{10}$ with
respect to $D_9$\ubmath}
\label{deddec}

In this appendix we present the first few complete levels of the
decomposition of $DE_{10}$ into representations of its $D_9$
subalgebra indicated in figure \ref{deddynk}. From the embedding
$DE_{10}\subset E_{10}$ it is clear that level
$\ell$ here has to be compared with representations on level
$2\ell$ in the preceding appendix \ref{eddec}.

{\small
\begin{longtable}{|c|c|c|c|c|c|r|}
\hline
$\ell$&\mbox{Dynkin $[p_s]$}&$\alpha\in DE_{10}$&$\alpha^2$&\mbox{mult}$(\alpha)$&$\mu$&\mbox{dim}
\\
\hline\hline
\endhead
1&[001000000]&(0,0,0,0,0,0,0,0,0,1)&2&1&1&816\\
1&[100000000]&(0,1,2,2,2,2,2,1,1,1)&0&8&0&18\\
\hline
2&[000001000]&(1,2,3,2,1,0,0,0,0,2)&2&1&1&18564\\
2&[101000000]&(0,1,2,2,2,2,2,1,1,2)&2&1&1&11475\\
2&[000100000]&(1,2,3,2,2,2,2,1,1,2)&0&8&0&3060\\
2&[200000000]&(0,2,4,4,4,4,4,2,2,2)&0&8&0&170\\
2&[010000000]&(1,2,4,4,4,4,4,2,2,2)&-2&44&1&153\\
2&[000000000]&(2,4,6,6,6,6,6,3,3,2)&-4&192&0&1\\
\hline
3&[010010000]&(1,2,4,3,2,2,2,1,1,3)&2&1&1&930240\\
3&[100000011]&(1,3,5,4,3,2,1,0,0,3)&2&1&1&707200\\
3&[100001000]&(1,3,5,4,3,2,2,1,1,3)&0&8&1&293760\\
3&[201000000]&(0,2,4,4,4,4,4,2,2,3)&2&1&1&90288\\
3&[000000020]&(2,4,6,5,4,3,2,0,1,3)&0&8&0&24310\\
3&[000000002]&(2,4,6,5,4,3,2,1,0,3)&0&8&0&24310\\
3&[011000000]&(1,2,4,4,4,4,4,2,2,3)&0&8&1&67830\\
3&[000000100]&(2,4,6,5,4,3,2,1,1,3)&-2&43&1&31824\\
3&[100100000]&(1,3,5,4,4,4,4,2,2,3)&-2&44&2&45696\\
3&[000010000]&(2,4,6,5,4,4,4,2,2,3)&-4&188&1&8568\\
3&[300000000]&(0,3,6,6,6,6,6,3,3,3)&0&8&0&1122\\
3&[110000000]&(1,3,6,6,6,6,6,3,3,3)&-4&192&2&1920\\
3&[001000000]&(2,4,6,6,6,6,6,3,3,3)&-6&711&2&816\\
3&[100000000]&(2,5,8,8,8,8,8,4,4,3)&-8&2408&1&18\\
\hline
4&[000101000]&(2,4,6,4,3,2,2,1,1,4)&2&1&1&26686260\\
4&[110000100]&(1,3,6,5,4,3,2,1,1,4)&2&1&1&43084800\\
4&[001000020]&(2,4,6,5,4,3,2,0,1,4)&2&1&1&13579566\\
4&[001000002]&(2,4,6,5,4,3,2,1,0,4)&2&1&1&13579566\\
4&[001000100]&(2,4,6,5,4,3,2,1,1,4)&0&8&1&17005950\\
4&[101100000]&(1,3,5,4,4,4,4,2,2,4)&2&1&1&13590225\\
4&[110010000]&(1,3,6,5,4,4,4,2,2,4)&0&8&2&10340352\\
4&[000200000]&(2,4,6,4,4,4,4,2,2,4)&0&8&0&2174436\\
4&[200000011]&(1,4,7,6,5,4,3,1,1,4)&0&8&2&6096948\\
4&[001010000]&(2,4,6,5,4,4,4,2,2,4)&-2&44&3&3837240\\
4&[010000011]&(2,4,7,6,5,4,3,1,1,4)&-2&43&3&5290740\\
4&[200001000]&(1,4,7,6,5,4,4,2,2,4)&-2&44&3&2494206\\
4&[010001000]&(2,4,7,6,5,4,4,2,2,4)&-4&188&4&2131800\\
4&[030000000]&(1,2,6,6,6,6,6,3,3,4)&2&1&1&261800\\
4&[301000000]&(0,3,6,6,6,6,6,3,3,4)&2&1&1&516800\\
4&[100000020]&(2,5,8,7,6,5,4,1,2,4)&-4&184&2&393822\\
4&[100000002]&(2,5,8,7,6,5,4,2,1,4)&-4&184&2&393822\\
4&[111000000]&(1,3,6,6,6,6,6,3,3,4)&-2&44&3&709632\\
4&[100000100]&(2,5,8,7,6,5,4,2,2,4)&-6&699&7&510510\\
4&[200100000]&(1,4,7,6,6,6,6,3,3,4)&-4&192&4&373065\\
4&[002000000]&(2,4,6,6,6,6,6,3,3,4)&-4&192&2&188955\\
4&[010100000]&(2,4,7,6,6,6,6,3,3,4)&-6&711&8&302328\\
4&[100010000]&(2,5,8,7,6,6,6,3,3,4)&-8&2376&8&132600\\
4&[000000011]&(3,6,9,8,7,6,5,2,2,4)&-8&2335&4&43758\\
4&[000001000]&(3,6,9,8,7,6,6,3,3,4)&-10&7317&7&18564\\
4&[400000000]&(0,4,8,8,8,8,8,4,4,4)&0&8&0&5814\\
4&[210000000]&(1,4,8,8,8,8,8,4,4,4)&-6&726&5&14212\\
4&[020000000]&(2,4,8,8,8,8,8,4,4,4)&-8&2408&3&8550\\
4&[101000000]&(2,5,8,8,8,8,8,4,4,4)&-10&7426&11&11475\\
4&[000100000]&(3,6,9,8,8,8,8,4,4,4)&-12&21394&6&3060\\
4&[200000000]&(2,6,10,10,10,10,10,5,5,4)&-12&21680&3&170\\
4&[010000000]&(3,6,10,10,10,10,10,5,5,4)&-14&59095&8&153\\
4&[000000000]&(4,8,12,12,12,12,12,6,6,4)&-16&155514&0&1\\
\hline
5&[100010100]&(2,5,8,6,4,3,2,1,1,5)&2&1&1&1522105200\\
5&[011000011]&(2,4,7,6,5,4,3,1,1,5)&2&1&1&1655413760\\
5&[010110000]&(2,4,7,5,4,4,4,2,2,5)&2&1&1&748261800\\
5&[100100011]&(2,5,8,6,5,4,3,1,1,5)&0&8&3&1024287264\\
5&[201001000]&(1,4,7,6,5,4,4,2,2,5)&2&1&1&848300544\\
5&[000001002]&(3,6,9,7,5,3,2,1,0,5)&2&1&1&150760896\\
5&[000001020]&(3,6,9,7,5,3,2,0,1,5)&2&1&1&150760896\\
5&[011001000]&(2,4,7,6,5,4,4,2,2,5)&0&8&2&613958400\\
5&[000001100]&(3,6,9,7,5,3,2,1,1,5)&0&8&1&154187280\\
5&[100020000]&(2,5,8,6,4,4,4,2,2,5)&0&8&1&195699240\\
5&[100101000]&(2,5,8,6,5,4,4,2,2,5)&-2&44&5&359165664\\
5&[210000002]&(1,4,8,7,6,5,4,2,1,5)&2&1&1&232792560\\
5&[210000020]&(1,4,8,7,6,5,4,1,2,5)&2&1&1&232792560\\
5&[020000002]&(2,4,8,7,6,5,4,2,1,5)&0&8&1&136334016\\
5&[020000020]&(2,4,8,7,6,5,4,1,2,5)&0&8&1&136334016\\
5&[000010011]&(3,6,9,7,5,4,3,1,1,5)&-2&43&3&164466432\\
5&[210000100]&(1,4,8,7,6,5,4,2,2,5)&0&8&2&293862816\\
5&[101000002]&(2,5,8,7,6,5,4,2,1,5)&-2&43&4&174888350\\
5&[101000020]&(2,5,8,7,6,5,4,1,2,5)&-2&43&4&174888350\\
5&[020000100]&(2,4,8,7,6,5,4,2,2,5)&-2&43&4&171348100\\
5&[101000100]&(2,5,8,7,6,5,4,2,2,5)&-4&188&10&217443600\\
5&[120100000]&(1,3,7,6,6,6,6,3,3,5)&2&1&1&106419456\\
5&[003000000]&(2,4,6,6,6,6,6,3,3,5)&2&1&1&19969950\\
5&[000011000]&(3,6,9,7,5,4,4,2,2,5)&-4&188&4&50605056\\
5&[201100000]&(1,4,7,6,6,6,6,3,3,5)&0&8&2&95178240\\
5&[000100020]&(3,6,9,7,6,5,4,1,2,5)&-4&184&4&42401502\\
5&[011100000]&(2,4,7,6,6,6,6,3,3,5)&-2&44&4&66853248\\
5&[000100002]&(3,6,9,7,6,5,4,2,1,5)&-4&184&4&42401502\\
5&[000100100]&(3,6,9,7,6,5,4,2,2,5)&-6&699&11&51395760\\
5&[210010000]&(1,4,8,7,6,6,6,3,3,5)&-2&44&5&69227298\\
5&[020010000]&(2,4,8,7,6,6,6,3,3,5)&-4&188&6&39673920\\
5&[100200000]&(2,5,8,6,6,6,6,3,3,5)&-4&192&5&28139760\\
5&[300000011]&(1,5,9,8,7,6,5,2,2,5)&-2&44&4&37209600\\
5&[101010000]&(2,5,8,7,6,6,6,3,3,5)&-6&711&17&47837592\\
5&[110000011]&(2,5,9,8,7,6,5,2,2,5)&-6&699&19&60648588\\
5&[001000011]&(3,6,9,8,7,6,5,2,2,5)&-8&2335&20&24186240\\
5&[300001000]&(1,5,9,8,7,6,6,3,3,5)&-4&192&6&15049216\\
5&[110001000]&(2,5,9,8,7,6,6,3,3,5)&-8&2376&26&24066900\\
5&[000110000]&(3,6,9,7,6,6,6,3,3,5)&-8&2376&12&9883800\\
5&[001001000]&(3,6,9,8,7,6,6,3,3,5)&-10&7317&28&9302400\\
5&[130000000]&(1,3,8,8,8,8,8,4,4,5)&0&8&1&2154240\\
5&[401000000]&(0,4,8,8,8,8,8,4,4,5)&2&1&1&2386800\\
5&[200000002]&(2,6,10,9,8,7,6,3,2,5)&-8&2344&11&3401190\\
5&[200000020]&(2,6,10,9,8,7,6,2,3,5)&-8&2344&11&3401190\\
5&[211000000]&(1,4,8,8,8,8,8,4,4,5)&-4&192&7&4558176\\
5&[021000000]&(2,4,8,8,8,8,8,4,4,5)&-6&711&9&2474010\\
5&[200000100]&(2,6,10,9,8,7,6,3,3,5)&-10&7340&27&4377296\\
5&[010000002]&(3,6,10,9,8,7,6,3,2,5)&-10&7204&18&2956096\\
5&[010000020]&(3,6,10,9,8,7,6,2,3,5)&-10&7204&18&2956096\\
5&[102000000]&(2,5,8,8,8,8,8,4,4,5)&-8&2408&13&2217072\\
5&[300100000]&(1,5,9,8,8,8,8,4,4,5)&-6&726&9&2188800\\
5&[010000100]&(3,6,10,9,8,7,6,3,3,5)&-12&21121&40&3779100\\
5&[110100000]&(2,5,9,8,8,8,8,4,4,5)&-10&7426&36&3281850\\
5&[001100000]&(3,6,9,8,8,8,8,4,4,5)&-12&21394&29&1116288\\
5&[200010000]&(2,6,10,9,8,8,8,4,4,5)&-12&21472&34&1108536\\
5&[010010000]&(3,6,10,9,8,8,8,4,4,5)&-14&58468&54&930240\\
5&[100000011]&(3,7,11,10,9,8,7,3,3,5)&-14&57786&47&707200\\
5&[100001000]&(3,7,11,10,9,8,8,4,4,5)&-16&152634&57&293760\\
5&[500000000]&(0,5,10,10,10,10,10,5,5,5)&0&8&0&25194\\
5&[310000000]&(1,5,10,10,10,10,10,5,5,5)&-8&2464&8&78336\\
5&[120000000]&(2,5,10,10,10,10,10,5,5,5)&-12&21680&20&84150\\
5&[201000000]&(2,6,10,10,10,10,10,5,5,5)&-14&59361&38&90288\\
5&[011000000]&(3,6,10,10,10,10,10,5,5,5)&-16&154174&47&67830\\
5&[000000002]&(4,8,12,11,10,9,8,4,3,5)&-16&150996&15&24310\\
5&[000000020]&(4,8,12,11,10,9,8,3,4,5)&-16&150996&15&24310\\
5&[000000100]&(4,8,12,11,10,9,8,4,4,5)&-18&382919&34&31824\\
5&[100100000]&(3,7,11,10,10,10,10,5,5,5)&-18&386560&67&45696\\
5&[000010000]&(4,8,12,11,10,10,10,5,5,5)&-20&935832&38&8568\\
5&[300000000]&(2,7,12,12,12,12,12,6,6,5)&-16&156416&12&1122\\
5&[110000000]&(3,7,12,12,12,12,12,6,6,5)&-20&943724&44&1920\\
5&[001000000]&(4,8,12,12,12,12,12,6,6,5)&-22&2213755&39&816\\
5&[100000000]&(4,9,14,14,14,14,14,7,7,5)&-24&5086640&16&18\\
\hline
\end{longtable}
}
\end{section}

\end{document}